\documentclass[aps,prl,twocolumn,superscriptaddress]{revtex4-1}  
\usepackage{graphicx}  
\usepackage{dcolumn}   
\usepackage{bm}        
\usepackage{amssymb}   
\usepackage[final]{epsfig}
\usepackage{amssymb}
\usepackage{latexsym}
\usepackage{wrapfig}
\usepackage{amsmath}
\usepackage{booktabs}
\usepackage{threeparttable}
\usepackage{xcolor}
\usepackage{amsthm}
\usepackage{lipsum}
\usepackage{hyperref}
\usepackage{titlesec,enumitem}
\newcommand{\beq}{\begin{equation}}
\newcommand{\eeq}{\end{equation}}
\newcommand{\beqs}{\begin{eqnarray}}
\newcommand{\eeqs}{\end{eqnarray}}

\begin{document}



\title{Continuum field theory for the deformations of planar kirigami}
%
\author{Yue Zheng}
\affiliation{Aerospace and Mechanical Engineering, University of Southern California, Los Angeles, CA 90014, USA}
\author{Imtiar Niloy}%
\affiliation{Civil Engineering, Stony Brook University, Stony Brook, NY 11794, USA}
\author{Paolo Celli}%
\affiliation{Civil Engineering, Stony Brook University, Stony Brook, NY 11794, USA}
\author{Ian Tobasco}
\email{tobasco@uic.edu}
\affiliation{Mathematics, Statistics and Computer Science, University of Illinois at Chicago, Chicago, IL 60607, USA}
\author{Paul Plucinsky}
\email{plucinsk@usc.edu}
\affiliation{Aerospace and Mechanical Engineering, University of Southern California, Los Angeles, CA 90014, USA}%
\date{\today}

\begin{abstract}
Mechanical metamaterials exhibit exotic properties at the system level, that emerge from the interactions of many nearly rigid building blocks. Determining these emergent properties theoretically has remained an open challenge outside of a few select examples. 
Here, for a large class of periodic and planar kirigami, we provide a coarse-graining rule linking the design of the panels and slits to the kirigami's macroscale deformations. The procedure gives a system of nonlinear partial differential equations (PDE) expressing geometric compatibility of angle functions related to the motion of individual slits.  Leveraging known solutions of the PDE, we present excellent agreement between simulations and experiments across kirigami designs. The results reveal a surprising nonlinear wave response that persists even at large boundary loads, the existence of which is determined completely by the Poisson's ratio of the unit cell.
\end{abstract}

\maketitle
Mechanical metamaterials are solids with exotic 
properties arising primarily from the geometry and topology of their mesostructures.
Recent studies have focused on creating metamaterials with unexpected shape-morphing capabilities \cite{Mullin_PRL_2007, Bertoldi_NATREVMATS_2017}, as this property is advantageous in applications spanning robotics, bio-medical devices, and space structures \cite{rafsanjani2019programming,kuribayashi2006self,velvaluri2021origami,zirbel2013accommodating}. A natural motif in this setting is a design that exhibits a mechanism~\cite{Pellegrino_IJSS_1986, Hutchinson2006,milton2013complete} or floppy mode~\cite{Lubensky2015}: the pattern, when idealized as an assembly of rigid elements connected along perfect hinges, can be activated by a continuous motion at zero energy. Yet mechanisms, even when carefully designed, rarely occur as a natural response to loads \cite{Coulais2018}. Instead, the complex elastic interplay of a metamaterial's building blocks results in an exotic soft mode of deformation. Characterizing soft modes is a difficult problem. 
Linear analysis hints at a rich field theory \cite{alibert2003truss,abdoul2018strain}, the nonlinear version of which has been uncovered only in a few  examples. 
Miura-Origami \cite{Schenk2013}, for instance, takes on a saddle like shape under bending, a feature linked to its auxetic behavior in the plane \cite{Wei2013}. The Rotating Squares (RS) \cite{grima2000auxetic} pattern exhibits domain wall motion \cite{Deng2020} and was recently linked to conformal soft modes \cite{czajkowski2022conformal}. 

In this Letter, we go far beyond any one example to establish a general coarse-graining rule determining the exotic, nonlinear soft modes of a large class of mechanism-based mechanical metamaterials inspired by kirigami. Our method includes the RS pattern as a special case, illuminating the particular nature of its conformal response. In general, we find a dichotomy between kirigami systems that respond by a nonlinear wave-like motion, and others including conformal kirigami that do not. We turn to introduce the specific systems treated here, and to describe our theoretical and experimental results. 

\begin{figure}
\centering
\includegraphics[scale = 1]{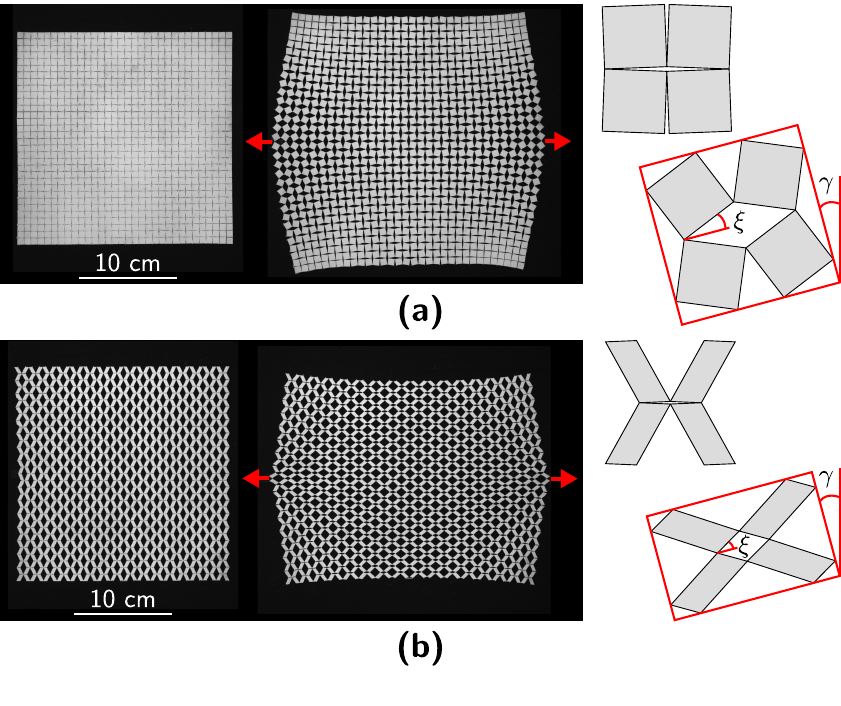}
\caption{Response of planar kirigami to the heterogeneous loading conditions indicated by the arrows. (a) Rotating Squares pattern; (b) another pattern featuring rhombi slits. Insets depict representative unit cells in reference and deformed configurations, with actuation angles $\xi$ and $\gamma$.}
\label{fig:IntroFig}
\end{figure}

\textit{Setup and overview of results\;--}\;Kirigami traditionally describes an elastic sheet with a pattern of cuts and folds \cite{Callens2017, Sussman_PNAS_2015, Wang2017}. More recently, the term has come to include cut patterns that, by themselves, produce complex deformations both in and out-of-plane \cite{ Cho_PNAS_2014, Rafsanjani2016, Tang2017, Blees2015, Rafsanjani2017, Dias_SOFTMATTER_2017, Konakovic2018, Celli2018, Choi2019}.\;Here, we study the 2D response of patterns with  repeating unit cells of four convex quadrilateral panels and four parallelogram slits. These patterns form a large model system for mechanism-based kirigami  \cite{Yang2018geometry, Singh2021,dang2022theorem}; their pure mechanism deformations are unit-cell periodic and counter-rotate the panels. Fig.\;\ref{fig:IntroFig} shows two examples, with the familiar RS pattern in (a). Each kirigami is free to deform as a mechanism under the loading, yet curiously neither does. Instead, exotic soft modes reveal themselves in the response.

What determines soft modes? The key insight is that each unit cell is  approximately mechanistic, yielding a bulk actuation that varies slowly from cell to cell. To characterize the response, then, one must solve the geometry problem of ``fitting together'' many nearly mechanistic cells. 
Coarse-graining this problem, we derive a continuum field theory coupling the kirigami's \textit{macroscopic} or \textit{effective deformation} to the local motion of its unit cells. For each cell, we track the opening angle  $2\xi$ of its deformed slit, along with an angle $\gamma$ giving the cell's overall rotation as in Fig.\;\ref{fig:IntroFig}. 
We derive a system of partial differential equations (PDEs) relating these angles, whose coefficients depend nonlinearly on $\xi$ as well as on the unit cell design. Solving this system exactly, we demonstrate an excellent match with experiments of different designs. 

Our theory divides  planar kirigami into two generic classes, which we term  \textit{elliptic} and \textit{hyperbolic} based on the so-called type of the coarse-grained PDE \cite{courant2008methods,evans10}. 
Elliptic kirigami shows a characteristic decay in actuation away from  loads. In contrast, hyperbolic kirigami deforms with persistent actuation, via a nonlinear wave response. Surprisingly, this dichotomy turns out to be directly related to the Poisson's ratio of the unit cell---elliptic kirigami is auxetic, while hyperbolic kirigami is not. This result serves as a powerful demonstration of our continuum field theory, and adds to the emerging literature connecting Poisson's ratio to the qualitative behavior of  mechanical metamaterials \cite{Wei2013, nassar2017curvature,lebee2018fitting,rocklin2017transformable}.




 
\begin{figure}
\centering
\includegraphics[scale=1]{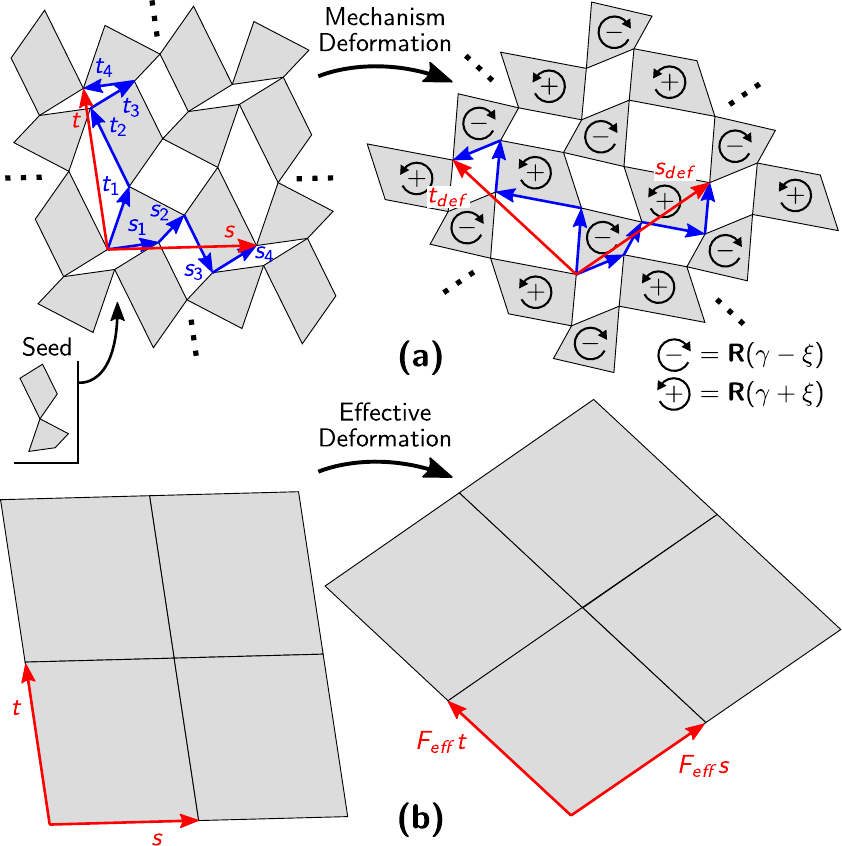}
\caption{Coarse-graining a mechanism. (a) Vectors $\mathbf{s}_i,\mathbf{t}_i$ define the unit cell, which tessellates along $\mathbf{s}$ and $\mathbf{t}$ to produce the pattern. In a mechanism, panels rotate by the indicated rotation matrices $\mathbf{R}(\gamma \pm \xi)$.  (b) Coarse-graining through the lattice defines the effective deformation gradient $\mathbf{F}_{\text{eff}}$. Soft modes agree locally with this picture.}
\label{fig:RigidDef}
\end{figure}

\textit{Coarse-graining planar kirigami\;--}\;We begin by introducing a general kirigami pattern consisting of a periodic array of unit cells, each having four quad panels and four parallelogram slits as in Fig.\;\ref{fig:RigidDef}(a). The most general setup is  as follows: start by selecting a seed of two quad panels connected at a corner point, rotate a copy of this seed $180^{\mathrm{o}}$, and connect it to the original seed to form a unit cell. Provided the resulting panels are disjoint, tessellating this unit cell along a Bravais lattice with basis vectors $\mathbf{s} = \mathbf{s}_1 + \mathbf{s}_2 + \mathbf{s}_3 + \mathbf{s}_4$ and $\mathbf{t} = \mathbf{t}_1 + \mathbf{t}_2 + \mathbf{t}_3 + \mathbf{t}_4$ gives a viable pattern. For an explanation of why this procedure is exhaustive, see supplemental section SM.1~\cite{suppl}. We fix one such pattern  and coarse-grain its kinematics. 
 
First, we consider mechanisms. As our kirigami has parallelogram slits, its pure mechanism deformations are given by an alternating array of panel rotations specified by the  rotation matrices $\mathbf{R}(\gamma \pm\xi)$ in Fig.\;\ref{fig:RigidDef}(a). 
(The angles $\gamma$ and $\xi$ agree with those of Fig.~\ref{fig:IntroFig}.) 
To coarse-grain, we view the deformation as distorting the underlying Bravais lattice: from the top half of the figure, the original lattice vectors $\mathbf{s}$ and $\mathbf{t}$ deform to 
\begin{equation}
\begin{aligned}\label{eq:defBravais}
&\mathbf{s}_{\text{def}} = \mathbf{R}(\gamma) \big( \mathbf{R}(-\xi)  (\mathbf{s}_{1} + \mathbf{s}_2) + \mathbf{R}(\xi) ( \mathbf{s}_{3} + \mathbf{s}_4) \big),\\
&\mathbf{t}_{\text{def}} = \mathbf{R}(\gamma) \big( \mathbf{R}(-\xi)  (\mathbf{t}_{1} + \mathbf{t}_4) + \mathbf{R}(\xi)  (\mathbf{t}_{2} + \mathbf{t}_3) \big).
\end{aligned}
\end{equation}
In turn, this distortion can be encoded into the two-by-two matrix $\mathbf{F}_\text{eff}$ defined by $\mathbf{F}_{\text{eff}} \mathbf{s} = \mathbf{s}_{\text{def}}$ and $\mathbf{F}_{\text{eff}} \mathbf{t} = \mathbf{t}_{\text{def}}$, concretely linking Fig.\;\ref{fig:RigidDef}(a) and (b). We call $\mathbf{F}_\text{eff}$ the \textit{coarse-grained} or \textit{effective deformation gradient} associated with the mechanism. Evidently,
\begin{equation}
\begin{aligned}\label{eq:Feff}
\mathbf{F}_{\text{eff}} = \mathbf{R}(\gamma) \mathbf{A}(\xi)
\end{aligned}
\end{equation}
for a shape tensor $\mathbf{A}(\xi)$ that depends only on $\xi$ and on the vectors $\mathbf{s}_i$ and $\mathbf{t}_i$ defining the unit cell. This tensor will be made explicit in the examples to come (see SM.2~\cite{suppl} for the general formula).

Having coarse-grained the pattern's mechanisms, we now extend our viewpoint to its \textit{exotic soft modes of deformation}, whose elastic energy scaling is less than bulk. Specifically, we consider elastic effects accounting for the finite size and distortion of the inter-panel hinges, and show in SM.3~\cite{suppl} that the energy per unit area of the kirigami vanishes with an increasing number of cells provided its effective deformation $\mathbf{y}_{\text{eff}}(\mathbf{x})$ obeys 
\begin{equation}
\begin{aligned}\label{eq:effectiveDescription}
\nabla \mathbf{y}_{\text{eff}}(\mathbf{x}) = \mathbf{R}(\gamma(\mathbf{x})) \mathbf{A}(\xi(\mathbf{x})).
\end{aligned}
\end{equation} 
While this PDE is trivially solved by the pure mechanisms in (\ref{eq:Feff}), it admits many other exotic solutions whose effective deformation gradients $\nabla\mathbf{y}_\text{eff}(\mathbf{x})$ and angle fields $\gamma(\mathbf{x})$ and $\xi(\mathbf{x})$ vary across the sample. 
The PDE characterizes soft modes in a doubly asymptotic limit of finely patterned kirigami, in which the hinges are small relative to the panels and the number of panels is large.

As gradients are curl-free, it follows by taking the curl of (\ref{eq:effectiveDescription}) that (SM.4~\cite{suppl})
\begin{equation}
\begin{aligned}\label{eq:compat} 
 \nabla \gamma(\mathbf{x}) = \boldsymbol{\Gamma}(\xi(\mathbf{x})) \nabla \xi(\mathbf{x}) 
 \end{aligned}
 \end{equation}
 for  $\boldsymbol{\Gamma}(\xi) =  \frac{\mathbf{A}^T(\xi)\mathbf{A}'(\xi) }{\det \mathbf{A}(\xi)}  \mathbf{R}(\tfrac{\pi}{2})$.
Eq.\;(\ref{eq:compat}) is a PDE reflecting the  geometric constraint that every closed loop in the kirigami must remain closed. This PDE can sometimes be solved analytically for the angle fields, as we do in the examples below, but in general we imagine it will be solved numerically. After finding $\gamma(\mathbf{x})$ and $\xi(\mathbf{x})$,  $\mathbf{y}_{\text{eff}}(\mathbf{x})$ can be recovered from  (\ref{eq:effectiveDescription})  uniquely up to a translation.  Eqs.\;(\ref{eq:effectiveDescription}-\ref{eq:compat}) furnish a complete  effective description of the locally mechanistic kinematics of any planar kirigami with a unit cell of four quad panels and four parallelogram slits.



\textit{Linear analysis, PDE type and Poisson's ratio\;--} 
While the effective description (\ref{eq:effectiveDescription}-\ref{eq:compat}) is nonlinear, we can start to learn its implications for kirigami soft modes by linearizing about a pure mechanism. 
We do so first for the class of rhombi-slit kirigami, whose shape tensors $\mathbf{A}(\xi)$ are diagonal. This simplification greatly clarifies the exposition without compromising the generality of our results; we treat general patterns at the end of this section. 

\begin{figure}
\centering
\includegraphics[scale = 1]{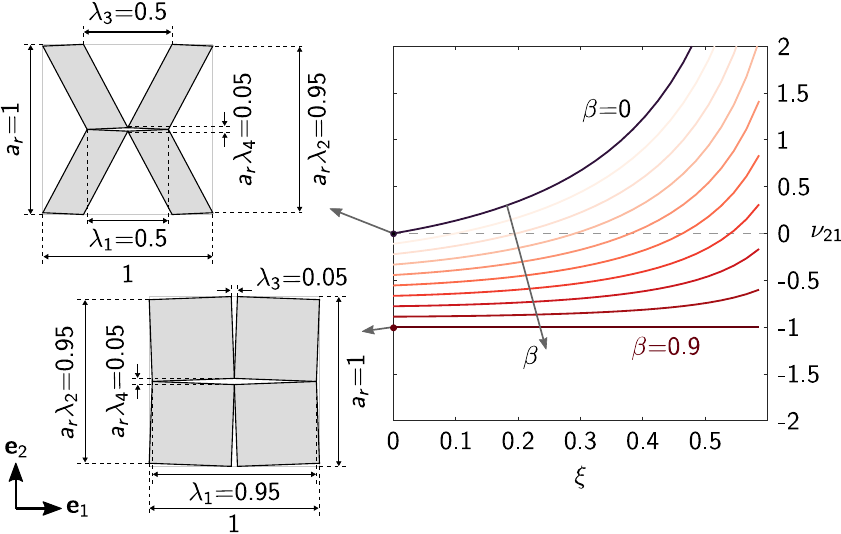}
\caption{Effective Poisson's ratio as a function of slit actuation $\xi$ for different rhombi-slit kirigami. The plot fixes $\alpha = -0.9$ and varies $\beta$ from $0$ to $0.9$. The RS pattern on the lower left sits at the lower extreme $\beta = 0.9$. It is purely dilational ($\nu_{21} = -1$) and is auxetic for all $\xi$. The upper extreme $\beta=0$ arises with the design on the upper left, which is non-auxetic $(\nu_{21}>0)$ for all relevant $\xi >0$. Some designs transition between auxetic and non-auxetic behavior as a function of $\xi$.}
\label{fig:PRatio}
\end{figure}

 
Per Fig.\;\ref{fig:PRatio}, a rhombi-slit kirigami is defined by parameters  $\lambda_{1}, \ldots, \lambda_4$ that can take any value in $[0,1]$, and an aspect ratio $a_r >0$: 
\begin{equation}
\begin{aligned}\label{eq:rhombiSlits}
&\mathbf{A}(\xi) = \mu_1(\xi) \mathbf{e}_1 \otimes \mathbf{e}_1 + \mu_2(\xi) \mathbf{e}_2 \otimes \mathbf{e}_2, \\
&\mu_1(\xi) = \cos \xi - \alpha \sin \xi,\quad \mu_2(\xi)  = \cos \xi + \beta \sin \xi, \\
&\alpha = a_r (\lambda_4 - \lambda_2), \quad\beta =  a_r^{-1} (\lambda_1 - \lambda_3).
\end{aligned}
\end{equation}
Note $\alpha$ and $\beta$ encode the geometry of the unit cell, $\mu_{1}(\xi)$ and $\mu_2(\xi)$ give the stretch or contraction of its sides under a mechanism, and  $\mathbf{e}_{1}$ and $\mathbf{e}_2$ are unit vectors along the initial slit axes, as in Fig.\;\ref{fig:PRatio}. Finally,  $\boldsymbol{\Gamma}(\xi)$ in (\ref{eq:compat}) satisfies 
\begin{equation}
    \begin{aligned}\label{eq:Gamma}
    \boldsymbol{\Gamma}(\xi) = \Gamma_{12}(\xi) \mathbf{e}_1 \otimes \mathbf{e}_2 + \Gamma_{21}(\xi) \mathbf{e}_2 \otimes \mathbf{e}_1
    \end{aligned}
\end{equation} 
for $\Gamma_{12}(\xi)  = -\mu_1'(\xi)/\mu_{2}(\xi)$ and $\Gamma_{21}(\xi) = \mu_2'(\xi)/ \mu_1(\xi)$. 
Eqs.\;(\ref{eq:rhombiSlits}-\ref{eq:Gamma}) follow from (\ref{eq:defBravais}-\ref{eq:Feff}) after choosing appropriate $\mathbf{s}_i$ and $\mathbf{t}_i$ (SM.2~\cite{suppl}).

Proceeding perturbatively, we write $\xi(\mathbf{x})=\xi_0+\delta\xi(\mathbf{x})$ and $\gamma(\mathbf{x})=\delta\gamma(\mathbf{x})$ for small angles $\delta \xi(\mathbf{x})$ and $\delta \gamma (\mathbf{x})$, and let $\mathbf{y}_{\text{eff}}(\mathbf{x}) = \mathbf{A}(\xi_0)\mathbf{x} + \mathbf{u}(\mathbf{A}(\xi_0)\mathbf{x})$ for a small  displacement $\mathbf{u}(\mathbf{y})$ about a pure mechanism with constant slit opening angle $2\xi_0$. 
(Taking $\gamma_0=0$ eliminates a free global rotation.) 
Expanding (\ref{eq:effectiveDescription}) to linear order and computing the strain $\boldsymbol{\varepsilon}(\mathbf{y}) =\tfrac{1}{2} (\nabla \mathbf{u}(\mathbf{y}) + \nabla \mathbf{u}^T(\mathbf{y}))$ yields
\begin{equation}
\begin{aligned}\label{eq:strain}
&\boldsymbol{\varepsilon}(\mathbf{A}(\xi_0)\mathbf{x}) = \delta \xi(\mathbf{x}) \left(\begin{array}{cc} \varepsilon_1(\xi_0) & 0 \\ 0 & \varepsilon_2(\xi_0) \end{array}\right)
\end{aligned}
\end{equation}
with $\varepsilon_i(\xi_0) = \mu_i'(\xi_0)/\mu_i(\xi_0)$, $i=1,2$. Similarly, expanding (\ref{eq:compat}) to linear order and taking its curl gives that
\begin{equation}
    \begin{aligned}\label{eq:linPDE}
    0=\big(\Gamma_{21}(\xi_{0})\partial_{1}^{2} -\Gamma_{12}(\xi_{0})\partial_{2}^{2}\big)\delta\xi(\mathbf{x}).
    \end{aligned}
\end{equation}
Both equations must hold for the  perturbation to be consistent with the effective theory. 

 \begin{figure*}
\centering
\includegraphics[scale = 1]{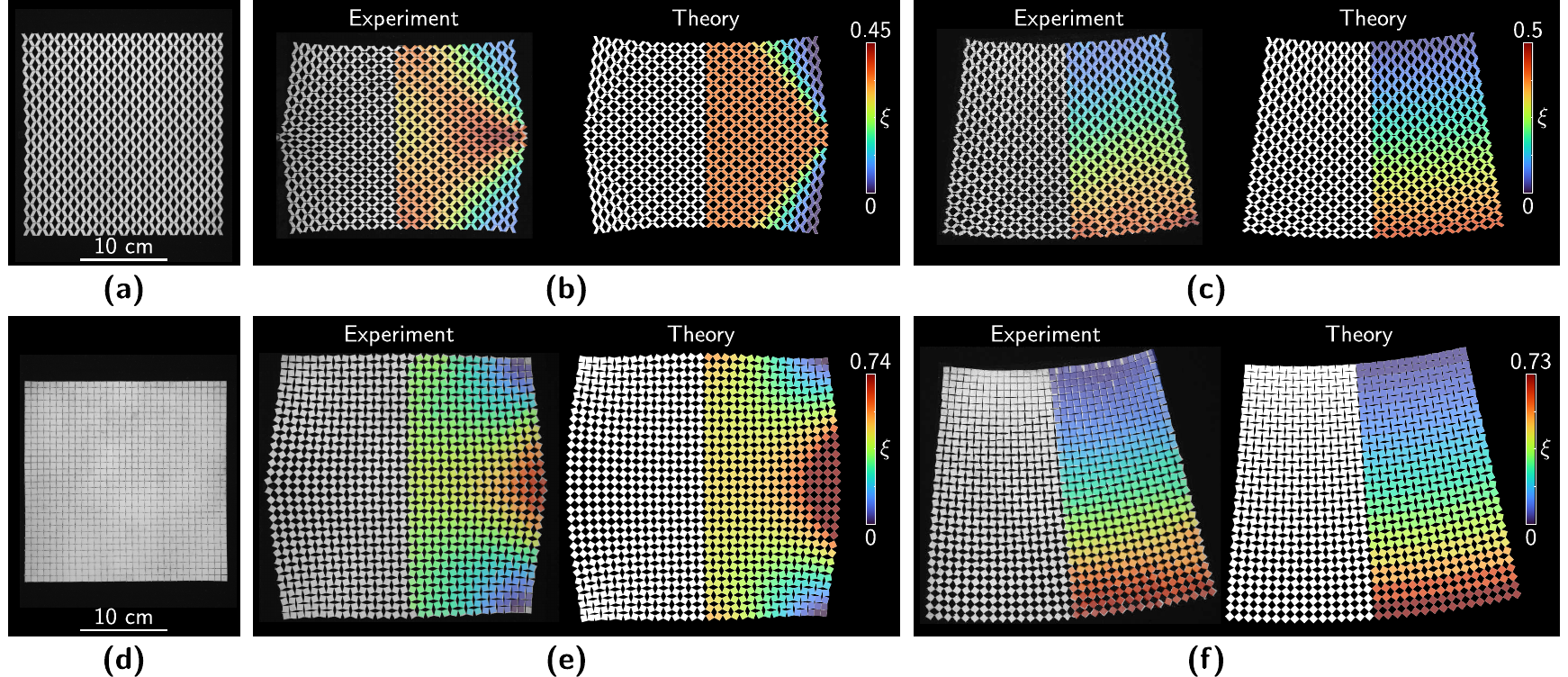}
\caption{Comparison between theory and experiments of rhombi-slit kirigami. (a,d) Two $16\times16$ cell patterns prior to deformation, with opposite Poisson's ratios and types. Top row is non-auxetic and hyperbolic. Bottom row is auxetic and elliptic. (b,e) Left entries are experimental samples pulled along their centerlines. Right entries show simulations, based on exact solutions of the effective theory. (c,f) Annular deformations produced experimentally (left) and using the theory (right). Colormaps indicate the slit actuation angle $\xi(\mathbf{x})$, extracted from the experiment using the procedure in SM.7~\cite{suppl}.}
\label{fig:FinalFig}
\end{figure*} 

The ratio of principal strains in (\ref{eq:strain}) defines an \textit{effective Poisson's ratio} which turns out to be  directly related to the coefficients in (\ref{eq:linPDE}):
\begin{equation}\label{eq:firstPoissons}
    \begin{aligned}
    \nu_{21}(\xi_0) := -\frac{\varepsilon_2(\xi_0)}{\varepsilon_1(\xi_0)} = \frac{\Gamma_{21}(\xi_{0})}{\Gamma_{12}(\xi_0)}\frac{\mu_1^2(\xi_0)}{\mu_2^2(\xi_0)}.
    \end{aligned}
\end{equation}
This link has remarkable implications.\;Writing  (\ref{eq:linPDE})  as $\partial_2^2 \delta \xi(\mathbf{x}) = \tfrac{\mu_2^2(\xi_0)}{\mu_1^2(\xi_0)}\nu_{21}(\xi_0) \partial_1^2 \delta \xi(\mathbf{x})$ and applying standard PDE theory, we discover that the overall structure of the perturbations is governed by the sign of the Poisson's ratio, i.e., by whether the pattern is auxetic or not:
\begin{equation}
    \begin{aligned}
    \begin{cases}\label{eq:Poissons}
    \nu_{21}(\xi_0) < 0 \quad \text{ elliptic and auxetic},   \\
    \nu_{21}(\xi_0) > 0 \quad  \text{ hyperbolic and non-auxetic}.
    \end{cases}
    \end{aligned}
\end{equation}
This criterion is visualized in Fig.~3.


The terms hyperbolic and elliptic come from PDE theory where an equation's type, found by linearization, informs the structure of its solutions \cite{courant2008methods,evans10}. Here in the hyperbolic case,  (\ref{eq:linPDE}) is the classical wave
equation with wave speed $c=\tfrac{\mu_2(\xi_0)}{\mu_1(\xi_0)}\sqrt{\nu_{21}(\xi_0)}$, the $x_{1}$- and $x_{2}$-coordinates being like ``space'' and ``time''. Linearization predicts waves for small loads; motivated by this, we go on below to construct a branch of nonlinear wave solutions describing the hyperbolic kirigami in Fig.\;\ref{fig:IntroFig}(b).
In contrast, the RS pattern in Fig.\;\ref{fig:IntroFig}(a) is auxetic and so is elliptic.  Instead of waves, elliptic kirigami shows a decay in  actuation away from  loads. We highlight the strong maximum principle of elliptic PDEs \cite{evans10}: the maximum and minimum actuation in an elliptic kirigami must occur only at its boundary, lest it deform by a constant mechanism. No such principle holds for hyperbolic kirigami. 
 
Remarkably, the same coupling in (\ref{eq:Poissons}) between Poisson's ratio and PDE type holds for the general quad-based kirigami patterns treated here. 
 We sketch the main ideas to provide clarity on this important result (see SM.5~\cite{suppl} for details). 
 Linearizing about a mechanism as before leads to a strain   $\boldsymbol{\varepsilon}(\mathbf{A}(\xi_0)\mathbf{x})$ with eigenvalues $\delta \xi(\mathbf{x}) \varepsilon_i(\xi_0)$, $i=1,2$. 
 Passing to a principle frame, the effective Poisson's ratio  of the  pattern---which dictates its auxeticity---is still given by the first expression in (\ref{eq:firstPoissons}).
 Eq.\;(\ref{eq:linPDE}) becomes a general second order linear PDE $c_{ij}(\xi_0)\partial^2_{ij}\delta\xi(\mathbf{x}) =0$ (summation implied). It is elliptic or hyperbolic according to the sign of the discriminant of its coefficients. A coordinate transformation reveals (\ref{eq:Poissons}). 
 
\textit{Nonlinear analysis and examples\;--}\;The previous linear analysis addresses the character of the kirigami's response nearby a pure mechanism, but does not prescribe it at  finite loads. We now present several exact solutions of the nonlinear system (\ref{eq:effectiveDescription}-\ref{eq:compat}) which capture the deformations of the kirigami in Fig.\;\ref{fig:FinalFig}, far into the nonlinear response. Our solutions are based on known results from PDE theory, which we detail in SM.6~\cite{suppl} and summarize here. 
Using them, we simulate the panel motions with an ansatz that rotates and translates the panels to fit the solution. Due to the finiteness of the sample, one may expect slight deviations between theory and experiment, which scale with the relative panel size. See SM.3~\cite{suppl} for more details.   

\noindent\textit{(i)\;Nonlinear waves\;--\;}
 Fig.\;\ref{fig:FinalFig}(a)  shows the $\alpha=-0.9$, $\beta=0$ pattern from the top left of Fig.\;\ref{fig:PRatio}, which remains non-auxetic, thus hyperbolic, for $\xi \in  (0, 0.235 \pi)$.  The hyperbolicity is borne out through the existence of nonlinear \textit{simple wave solutions} to (\ref{eq:compat}),  defined by the criteria that $\xi=\xi(\theta(\mathbf{x}))$ and $\gamma = \gamma(\theta(\mathbf{x}))$ for a scalar function $\theta(\mathbf{x})$. As a result, the angles vary across envelopes of straight line segments called characteristic curves. The term ``simple wave'' comes from compressible gas dynamics, where the same functional form governs gas densities varying next to regions of constant density \cite{courant1999supersonic}. For kirigami, simple waves alleviate slit openings next to regions of uniform actuation in response to loads. 
 
The left part of Fig.\;\ref{fig:FinalFig}(b) shows the experimental specimen pulled at its left and right ends along its centerline.  Slits open by an essentially constant amount in a central diamond region (orange), and recede towards the specimen's corners. Note the  ``fanning out" of contours of constant slit actuation from where the loads are applied. A simulation based on simple wave solutions matches these features on the right of Fig.\;\ref{fig:FinalFig}(b). Its straight line contours are characteristic curves.

\noindent\textit{(ii)\;Conformal maps\;--\;}Recent work~\cite{czajkowski2022conformal} has noted the relevance of conformal maps for kirigami. Adding to this discussion, and as an example of our more general elliptic class, we note using  \eqref{eq:rhombiSlits} that the only rhombi-slit kirigami designs that deform conformally ($\mu_1(\xi)=\mu_2(\xi)$ for all $\xi$ by definition \cite{do2016differential}) have $\alpha=-\beta$ and $\nu_{21}(\xi) = -1$.  This includes the RS pattern in Fig.\;\ref{fig:FinalFig}(d), fabricated according to the lower left $\alpha = -0.9$ design in Fig.\;\ref{fig:PRatio}. We highlight the RS pattern due to its dramatic shape-morphing. Conformal mappings are basic examples in complex analysis \cite{brown2009complex}, enabling  numerous solutions to (\ref{eq:compat}).

The left part of Fig.\;\ref{fig:FinalFig}(e) shows the RS pattern pulled at its left and right ends. Its slits open up dramatically at the loading points and remain closed at the corners: the largest and smallest openings are at the boundary, per the maximum principle. Contours of constant slit actuation form arcs around these points. On the right of Fig.\;\ref{fig:FinalFig}(e), we fit the deformed boundary of the pattern to a conformal map. The simulation recovers the locations where the slits are most open and closed, and qualitatively matches their variations in the bulk. 

\noindent\textit{(iii)\;Annuli\;--\;}Though one may think of hyerperbolic and elliptic kirigami as a dichotomy, and this is true as far as auxeticity is concerned, we close by pointing out the existence of some special effective deformations that are ``universal" in that they occur for both.  One example is the annular deformation in Fig.\;\ref{fig:FinalFig}(c) and (f), which arises from  (\ref{eq:compat}) under the condition that $\xi(\mathbf{x})$ is either only a function of $x_1$ or of $x_2$.  All rhombi-slit kirigami patterns are capable of this deformation, as we demonstrate using the previous hyperbolic (c) and elliptic (f) designs. Note unlike the previous examples, these experiments are done using pure displacement boundary conditions.

\textit{Discussion --} Looking forward, while our emphasis here was on the derivation of  coarse-grained PDEs expressing bulk geometric constraints, we set aside the important question of the forces underlying them. Understanding the inter-panel forces more closely should eventually lead to a complete continuum theory predicting exactly which exotic soft mode will arise in response to a given load. Our results show that the effective PDE system (\ref{eq:effectiveDescription}-\ref{eq:compat}) plays the dominant, constraining role. This is consistent with the conformal elasticity of Ref.~\cite{czajkowski2022conformal}.

More broadly, we expect that an effective PDE of a geometric origin exists to constrain the bulk behavior of mechanical metamaterials beyond kirigami. Such PDEs have been found for certain origami designs \cite{nassar2017curvature,lebee2018fitting}, via a differential geometric argument akin to our passage from (\ref{eq:effectiveDescription}) to (\ref{eq:compat}). In origami, one also finds a surprising coupling between the Poisson's ratio of the mechanisms and certain fine features of  exotic soft modes. Are such couplings universal? What about the role of heterogeneity \cite{Dudte2016,Celli2018,Choi2019,dang2022inverse}? Can coarse-graining lead to constitutive models for mechanical metamaterials, common to practical engineering \cite{khajehtourian2021continuum,mcmahan2021effective}, or to effective descriptions of their dynamics~\cite{Deng2017}? While there are many avenues left to explore, our work on the soft modes of planar kirigami highlights new physics and is a convincing step towards the discovery of a continuum theory for mechanical metamaterials at large.

\textbf{Acknowledgment.} YZ and PP acknowledge support through PP's start-up package at University of Southern California. IN and PC acknowledge the support of the Research Foundation for the State University of New York, and thank Megan Kam of iCreate for fabrication support. IT was supported by NSF award DMS-2025000.

\bibliographystyle{apsrev4-1} 
\bibliography{bib}

\begin{thebibliography}{47}%
\makeatletter
\providecommand \@ifxundefined [1]{%
 \@ifx{#1\undefined}
}%
\providecommand \@ifnum [1]{%
 \ifnum #1\expandafter \@firstoftwo
 \else \expandafter \@secondoftwo
 \fi
}%
\providecommand \@ifx [1]{%
 \ifx #1\expandafter \@firstoftwo
 \else \expandafter \@secondoftwo
 \fi
}%
\providecommand \natexlab [1]{#1}%
\providecommand \enquote  [1]{``#1''}%
\providecommand \bibnamefont  [1]{#1}%
\providecommand \bibfnamefont [1]{#1}%
\providecommand \citenamefont [1]{#1}%
\providecommand \href@noop [0]{\@secondoftwo}%
\providecommand \href [0]{\begingroup \@sanitize@url \@href}%
\providecommand \@href[1]{\@@startlink{#1}\@@href}%
\providecommand \@@href[1]{\endgroup#1\@@endlink}%
\providecommand \@sanitize@url [0]{\catcode `\\12\catcode `\$12\catcode
  `\&12\catcode `\#12\catcode `\^12\catcode `\_12\catcode `\%12\relax}%
\providecommand \@@startlink[1]{}%
\providecommand \@@endlink[0]{}%
\providecommand \url  [0]{\begingroup\@sanitize@url \@url }%
\providecommand \@url [1]{\endgroup\@href {#1}{\urlprefix }}%
\providecommand \urlprefix  [0]{URL }%
\providecommand \Eprint [0]{\href }%
\providecommand \doibase [0]{http://dx.doi.org/}%
\providecommand \selectlanguage [0]{\@gobble}%
\providecommand \bibinfo  [0]{\@secondoftwo}%
\providecommand \bibfield  [0]{\@secondoftwo}%
\providecommand \translation [1]{[#1]}%
\providecommand \BibitemOpen [0]{}%
\providecommand \bibitemStop [0]{}%
\providecommand \bibitemNoStop [0]{.\EOS\space}%
\providecommand \EOS [0]{\spacefactor3000\relax}%
\providecommand \BibitemShut  [1]{\csname bibitem#1\endcsname}%
\let\auto@bib@innerbib\@empty
\bibitem [{\citenamefont {Mullin}\ \emph {et~al.}(2007)\citenamefont {Mullin},
  \citenamefont {Deschanel}, \citenamefont {Bertoldi},\ and\ \citenamefont
  {Boyce}}]{Mullin_PRL_2007}%
  \BibitemOpen
  \bibfield  {author} {\bibinfo {author} {\bibfnamefont {T.}~\bibnamefont
  {Mullin}}, \bibinfo {author} {\bibfnamefont {S.}~\bibnamefont {Deschanel}},
  \bibinfo {author} {\bibfnamefont {K.}~\bibnamefont {Bertoldi}}, \ and\
  \bibinfo {author} {\bibfnamefont {M.~C.}\ \bibnamefont {Boyce}},\ }\href
  {\doibase 10.1103/PhysRevLett.99.084301} {\bibfield  {journal} {\bibinfo
  {journal} {Phys. Rev. Lett.}\ }\textbf {\bibinfo {volume} {99}},\ \bibinfo
  {pages} {084301} (\bibinfo {year} {2007})}\BibitemShut {NoStop}%
\bibitem [{\citenamefont {Bertoldi}\ \emph {et~al.}(2017)\citenamefont
  {Bertoldi}, \citenamefont {Vitelli}, \citenamefont {Christensen},\ and\
  \citenamefont {{van Hecke}}}]{Bertoldi_NATREVMATS_2017}%
  \BibitemOpen
  \bibfield  {author} {\bibinfo {author} {\bibfnamefont {K.}~\bibnamefont
  {Bertoldi}}, \bibinfo {author} {\bibfnamefont {V.}~\bibnamefont {Vitelli}},
  \bibinfo {author} {\bibfnamefont {J.}~\bibnamefont {Christensen}}, \ and\
  \bibinfo {author} {\bibfnamefont {M.}~\bibnamefont {{van Hecke}}},\ }\href
  {\doibase 10.1038/natrevmats.2017.66} {\bibfield  {journal} {\bibinfo
  {journal} {Nat. Rev. Mater.}\ }\textbf {\bibinfo {volume} {2}},\ \bibinfo
  {pages} {17066} (\bibinfo {year} {2017})}\BibitemShut {NoStop}%
\bibitem [{\citenamefont {Rafsanjani}\ \emph {et~al.}(2019)\citenamefont
  {Rafsanjani}, \citenamefont {Bertoldi},\ and\ \citenamefont
  {Studart}}]{rafsanjani2019programming}%
  \BibitemOpen
  \bibfield  {author} {\bibinfo {author} {\bibfnamefont {A.}~\bibnamefont
  {Rafsanjani}}, \bibinfo {author} {\bibfnamefont {K.}~\bibnamefont
  {Bertoldi}}, \ and\ \bibinfo {author} {\bibfnamefont {A.~R.}\ \bibnamefont
  {Studart}},\ }\href {\doibase 10.1126/scirobotics.aav7874} {\bibfield
  {journal} {\bibinfo  {journal} {Sci. Robot.}\ }\textbf {\bibinfo {volume}
  {4}},\ \bibinfo {pages} {eaav7874} (\bibinfo {year} {2019})}\BibitemShut
  {NoStop}%
\bibitem [{\citenamefont {Kuribayashi}\ \emph {et~al.}(2006)\citenamefont
  {Kuribayashi}, \citenamefont {Tsuchiya}, \citenamefont {You}, \citenamefont
  {Tomus}, \citenamefont {Umemoto}, \citenamefont {Ito},\ and\ \citenamefont
  {Sasaki}}]{kuribayashi2006self}%
  \BibitemOpen
  \bibfield  {author} {\bibinfo {author} {\bibfnamefont {K.}~\bibnamefont
  {Kuribayashi}}, \bibinfo {author} {\bibfnamefont {K.}~\bibnamefont
  {Tsuchiya}}, \bibinfo {author} {\bibfnamefont {Z.}~\bibnamefont {You}},
  \bibinfo {author} {\bibfnamefont {D.}~\bibnamefont {Tomus}}, \bibinfo
  {author} {\bibfnamefont {M.}~\bibnamefont {Umemoto}}, \bibinfo {author}
  {\bibfnamefont {T.}~\bibnamefont {Ito}}, \ and\ \bibinfo {author}
  {\bibfnamefont {M.}~\bibnamefont {Sasaki}},\ }\href {\doibase
  10.1016/j.msea.2005.12.016} {\bibfield  {journal} {\bibinfo  {journal}
  {Mater. Sci. Eng. A}\ }\textbf {\bibinfo {volume} {419}},\ \bibinfo {pages}
  {131} (\bibinfo {year} {2006})}\BibitemShut {NoStop}%
\bibitem [{\citenamefont {Velvaluri}\ \emph {et~al.}(2021)\citenamefont
  {Velvaluri}, \citenamefont {Soor}, \citenamefont {Plucinsky}, \citenamefont
  {de~Miranda}, \citenamefont {James},\ and\ \citenamefont
  {Quandt}}]{velvaluri2021origami}%
  \BibitemOpen
  \bibfield  {author} {\bibinfo {author} {\bibfnamefont {P.}~\bibnamefont
  {Velvaluri}}, \bibinfo {author} {\bibfnamefont {A.}~\bibnamefont {Soor}},
  \bibinfo {author} {\bibfnamefont {P.}~\bibnamefont {Plucinsky}}, \bibinfo
  {author} {\bibfnamefont {R.~L.}\ \bibnamefont {de~Miranda}}, \bibinfo
  {author} {\bibfnamefont {R.~D.}\ \bibnamefont {James}}, \ and\ \bibinfo
  {author} {\bibfnamefont {E.}~\bibnamefont {Quandt}},\ }\href {\doibase
  10.1038/s41598-021-90217-3} {\bibfield  {journal} {\bibinfo  {journal} {Sci.
  Rep.}\ }\textbf {\bibinfo {volume} {11}},\ \bibinfo {pages} {1} (\bibinfo
  {year} {2021})}\BibitemShut {NoStop}%
\bibitem [{\citenamefont {Zirbel}\ \emph {et~al.}(2013)\citenamefont {Zirbel},
  \citenamefont {Lang}, \citenamefont {Thomson}, \citenamefont {Sigel},
  \citenamefont {Walkemeyer}, \citenamefont {Trease}, \citenamefont {Magleby},\
  and\ \citenamefont {Howell}}]{zirbel2013accommodating}%
  \BibitemOpen
  \bibfield  {author} {\bibinfo {author} {\bibfnamefont {S.~A.}\ \bibnamefont
  {Zirbel}}, \bibinfo {author} {\bibfnamefont {R.~J.}\ \bibnamefont {Lang}},
  \bibinfo {author} {\bibfnamefont {M.~W.}\ \bibnamefont {Thomson}}, \bibinfo
  {author} {\bibfnamefont {D.~A.}\ \bibnamefont {Sigel}}, \bibinfo {author}
  {\bibfnamefont {P.~E.}\ \bibnamefont {Walkemeyer}}, \bibinfo {author}
  {\bibfnamefont {B.~P.}\ \bibnamefont {Trease}}, \bibinfo {author}
  {\bibfnamefont {S.~P.}\ \bibnamefont {Magleby}}, \ and\ \bibinfo {author}
  {\bibfnamefont {L.~L.}\ \bibnamefont {Howell}},\ }\href {\doibase
  10.1115/1.4025372} {\bibfield  {journal} {\bibinfo  {journal} {J. Mech.
  Des.}\ }\textbf {\bibinfo {volume} {135}},\ \bibinfo {pages} {111005}
  (\bibinfo {year} {2013})}\BibitemShut {NoStop}%
\bibitem [{\citenamefont {Pellegrino}\ and\ \citenamefont
  {Calladine}(1986)}]{Pellegrino_IJSS_1986}%
  \BibitemOpen
  \bibfield  {author} {\bibinfo {author} {\bibfnamefont {S.}~\bibnamefont
  {Pellegrino}}\ and\ \bibinfo {author} {\bibfnamefont {C.~R.}\ \bibnamefont
  {Calladine}},\ }\href {\doibase 10.1016/0020-7683(86)90014-4} {\bibfield
  {journal} {\bibinfo  {journal} {Int. J. Solids Struct.}\ }\textbf {\bibinfo
  {volume} {22}},\ \bibinfo {pages} {409} (\bibinfo {year} {1986})}\BibitemShut
  {NoStop}%
\bibitem [{\citenamefont {Hutchinson}\ and\ \citenamefont
  {Fleck}(2006)}]{Hutchinson2006}%
  \BibitemOpen
  \bibfield  {author} {\bibinfo {author} {\bibfnamefont {R.~G.}\ \bibnamefont
  {Hutchinson}}\ and\ \bibinfo {author} {\bibfnamefont {N.~A.}\ \bibnamefont
  {Fleck}},\ }\href {\doibase 10.1016/j.jmps.2005.10.008} {\bibfield  {journal}
  {\bibinfo  {journal} {J. Mech. Phys. Solids}\ }\textbf {\bibinfo {volume}
  {54}},\ \bibinfo {pages} {756} (\bibinfo {year} {2006})}\BibitemShut
  {NoStop}%
\bibitem [{\citenamefont {Milton}(2013)}]{milton2013complete}%
  \BibitemOpen
  \bibfield  {author} {\bibinfo {author} {\bibfnamefont {G.~W.}\ \bibnamefont
  {Milton}},\ }\href@noop {} {\bibfield  {journal} {\bibinfo  {journal} {J.
  Mech. Phys. Solids}\ }\textbf {\bibinfo {volume} {61}},\ \bibinfo {pages}
  {1543} (\bibinfo {year} {2013})}\BibitemShut {NoStop}%
\bibitem [{\citenamefont {Lubensky}\ \emph {et~al.}(2015)\citenamefont
  {Lubensky}, \citenamefont {Kane}, \citenamefont {Mao}, \citenamefont
  {Souslov},\ and\ \citenamefont {Sun}}]{Lubensky2015}%
  \BibitemOpen
  \bibfield  {author} {\bibinfo {author} {\bibfnamefont {T.~C.}\ \bibnamefont
  {Lubensky}}, \bibinfo {author} {\bibfnamefont {C.~L.}\ \bibnamefont {Kane}},
  \bibinfo {author} {\bibfnamefont {X.}~\bibnamefont {Mao}}, \bibinfo {author}
  {\bibfnamefont {A.}~\bibnamefont {Souslov}}, \ and\ \bibinfo {author}
  {\bibfnamefont {K.}~\bibnamefont {Sun}},\ }\href {\doibase
  10.1088/0034-4885/78/7/073901} {\bibfield  {journal} {\bibinfo  {journal}
  {Rep. Prog. Phys.}\ }\textbf {\bibinfo {volume} {78}},\ \bibinfo {pages}
  {073901} (\bibinfo {year} {2015})}\BibitemShut {NoStop}%
\bibitem [{\citenamefont {Coulais}\ \emph {et~al.}(2018)\citenamefont
  {Coulais}, \citenamefont {Kettenis},\ and\ \citenamefont {{van
  Hecke}}}]{Coulais2018}%
  \BibitemOpen
  \bibfield  {author} {\bibinfo {author} {\bibfnamefont {C.}~\bibnamefont
  {Coulais}}, \bibinfo {author} {\bibfnamefont {C.}~\bibnamefont {Kettenis}}, \
  and\ \bibinfo {author} {\bibfnamefont {M.}~\bibnamefont {{van Hecke}}},\
  }\href {\doibase 10.1038/nphys4269} {\bibfield  {journal} {\bibinfo
  {journal} {Nat. Phys.}\ }\textbf {\bibinfo {volume} {14}},\ \bibinfo {pages}
  {40} (\bibinfo {year} {2018})}\BibitemShut {NoStop}%
\bibitem [{\citenamefont {Alibert}\ \emph {et~al.}(2003)\citenamefont
  {Alibert}, \citenamefont {Seppecher},\ and\ \citenamefont
  {Dell’Isola}}]{alibert2003truss}%
  \BibitemOpen
  \bibfield  {author} {\bibinfo {author} {\bibfnamefont {J.-J.}\ \bibnamefont
  {Alibert}}, \bibinfo {author} {\bibfnamefont {P.}~\bibnamefont {Seppecher}},
  \ and\ \bibinfo {author} {\bibfnamefont {F.}~\bibnamefont {Dell’Isola}},\
  }\href@noop {} {\bibfield  {journal} {\bibinfo  {journal} {Math. Mech.
  Solids}\ }\textbf {\bibinfo {volume} {8}},\ \bibinfo {pages} {51} (\bibinfo
  {year} {2003})}\BibitemShut {NoStop}%
\bibitem [{\citenamefont {Abdoul-Anziz}\ and\ \citenamefont
  {Seppecher}(2018)}]{abdoul2018strain}%
  \BibitemOpen
  \bibfield  {author} {\bibinfo {author} {\bibfnamefont {H.}~\bibnamefont
  {Abdoul-Anziz}}\ and\ \bibinfo {author} {\bibfnamefont {P.}~\bibnamefont
  {Seppecher}},\ }\href@noop {} {\bibfield  {journal} {\bibinfo  {journal}
  {Math. Mech. Complex Syst.}\ }\textbf {\bibinfo {volume} {6}},\ \bibinfo
  {pages} {213} (\bibinfo {year} {2018})}\BibitemShut {NoStop}%
\bibitem [{\citenamefont {Schenk}\ and\ \citenamefont
  {Guest}(2013)}]{Schenk2013}%
  \BibitemOpen
  \bibfield  {author} {\bibinfo {author} {\bibfnamefont {M.}~\bibnamefont
  {Schenk}}\ and\ \bibinfo {author} {\bibfnamefont {S.~D.}\ \bibnamefont
  {Guest}},\ }\href {\doibase 10.1073/pnas.1217998110} {\bibfield  {journal}
  {\bibinfo  {journal} {Proc. Natl. Acad. Sci. U.S.A.}\ }\textbf {\bibinfo
  {volume} {110}},\ \bibinfo {pages} {3276} (\bibinfo {year}
  {2013})}\BibitemShut {NoStop}%
\bibitem [{\citenamefont {Wei}\ \emph {et~al.}(2013)\citenamefont {Wei},
  \citenamefont {Guo}, \citenamefont {Dudte}, \citenamefont {Liang},\ and\
  \citenamefont {Mahadevan}}]{Wei2013}%
  \BibitemOpen
  \bibfield  {author} {\bibinfo {author} {\bibfnamefont {Z.~Y.}\ \bibnamefont
  {Wei}}, \bibinfo {author} {\bibfnamefont {Z.~V.}\ \bibnamefont {Guo}},
  \bibinfo {author} {\bibfnamefont {L.}~\bibnamefont {Dudte}}, \bibinfo
  {author} {\bibfnamefont {H.~Y.}\ \bibnamefont {Liang}}, \ and\ \bibinfo
  {author} {\bibfnamefont {L.}~\bibnamefont {Mahadevan}},\ }\href {\doibase
  10.1103/PhysRevLett.110.215501} {\bibfield  {journal} {\bibinfo  {journal}
  {Phys. Rev. Lett.}\ }\textbf {\bibinfo {volume} {110}},\ \bibinfo {pages}
  {215501} (\bibinfo {year} {2013})}\BibitemShut {NoStop}%
\bibitem [{\citenamefont {Grima}\ and\ \citenamefont
  {Evans}(2000)}]{grima2000auxetic}%
  \BibitemOpen
  \bibfield  {author} {\bibinfo {author} {\bibfnamefont {J.~N.}\ \bibnamefont
  {Grima}}\ and\ \bibinfo {author} {\bibfnamefont {K.~E.}\ \bibnamefont
  {Evans}},\ }\href {\doibase 10.1023/A:1006781224002} {\bibfield  {journal}
  {\bibinfo  {journal} {J. Mater. Sci. Lett.}\ }\textbf {\bibinfo {volume}
  {19}},\ \bibinfo {pages} {1563} (\bibinfo {year} {2000})}\BibitemShut
  {NoStop}%
\bibitem [{\citenamefont {Deng}\ \emph {et~al.}(2020)\citenamefont {Deng},
  \citenamefont {Yu}, \citenamefont {Forte}, \citenamefont {Tournat},\ and\
  \citenamefont {Bertoldi}}]{Deng2020}%
  \BibitemOpen
  \bibfield  {author} {\bibinfo {author} {\bibfnamefont {B.}~\bibnamefont
  {Deng}}, \bibinfo {author} {\bibfnamefont {S.}~\bibnamefont {Yu}}, \bibinfo
  {author} {\bibfnamefont {A.~E.}\ \bibnamefont {Forte}}, \bibinfo {author}
  {\bibfnamefont {V.}~\bibnamefont {Tournat}}, \ and\ \bibinfo {author}
  {\bibfnamefont {K.}~\bibnamefont {Bertoldi}},\ }\href {\doibase
  10.1073/pnas.2015847117} {\bibfield  {journal} {\bibinfo  {journal} {Proc.
  Natl. Acad. Sci. U.S.A.}\ }\textbf {\bibinfo {volume} {117}},\ \bibinfo
  {pages} {31002} (\bibinfo {year} {2020})}\BibitemShut {NoStop}%
\bibitem [{\citenamefont {Czajkowski}\ \emph {et~al.}(2022)\citenamefont
  {Czajkowski}, \citenamefont {Coulais}, \citenamefont {van Hecke},\ and\
  \citenamefont {Rocklin}}]{czajkowski2022conformal}%
  \BibitemOpen
  \bibfield  {author} {\bibinfo {author} {\bibfnamefont {M.}~\bibnamefont
  {Czajkowski}}, \bibinfo {author} {\bibfnamefont {C.}~\bibnamefont {Coulais}},
  \bibinfo {author} {\bibfnamefont {M.}~\bibnamefont {van Hecke}}, \ and\
  \bibinfo {author} {\bibfnamefont {D.~Z.}\ \bibnamefont {Rocklin}},\ }\href
  {\doibase 10.1038/s41467-021-27825-0} {\bibfield  {journal} {\bibinfo
  {journal} {Nat. Commun.}\ }\textbf {\bibinfo {volume} {13}},\ \bibinfo
  {pages} {211} (\bibinfo {year} {2022})}\BibitemShut {NoStop}%
\bibitem [{\citenamefont {Callens}\ and\ \citenamefont
  {Zadpoor}(2018)}]{Callens2017}%
  \BibitemOpen
  \bibfield  {author} {\bibinfo {author} {\bibfnamefont {S.~J.~P.}\
  \bibnamefont {Callens}}\ and\ \bibinfo {author} {\bibfnamefont {A.~A.}\
  \bibnamefont {Zadpoor}},\ }\href {\doibase 10.1016/j.mattod.2017.10.004}
  {\bibfield  {journal} {\bibinfo  {journal} {Mater. Today}\ }\textbf {\bibinfo
  {volume} {21}},\ \bibinfo {pages} {241} (\bibinfo {year} {2018})}\BibitemShut
  {NoStop}%
\bibitem [{\citenamefont {Sussman}\ \emph {et~al.}(2015)\citenamefont
  {Sussman}, \citenamefont {Cho}, \citenamefont {Castle}, \citenamefont {Gong},
  \citenamefont {Jung}, \citenamefont {Yang},\ and\ \citenamefont
  {Kamien}}]{Sussman_PNAS_2015}%
  \BibitemOpen
  \bibfield  {author} {\bibinfo {author} {\bibfnamefont {D.~M.}\ \bibnamefont
  {Sussman}}, \bibinfo {author} {\bibfnamefont {Y.}~\bibnamefont {Cho}},
  \bibinfo {author} {\bibfnamefont {T.}~\bibnamefont {Castle}}, \bibinfo
  {author} {\bibfnamefont {X.}~\bibnamefont {Gong}}, \bibinfo {author}
  {\bibfnamefont {E.}~\bibnamefont {Jung}}, \bibinfo {author} {\bibfnamefont
  {S.}~\bibnamefont {Yang}}, \ and\ \bibinfo {author} {\bibfnamefont {R.~D.}\
  \bibnamefont {Kamien}},\ }\href {\doibase 10.1073/pnas.1506048112} {\bibfield
   {journal} {\bibinfo  {journal} {Proc. Natl. Acad. Sci. U.S.A.}\ }\textbf
  {\bibinfo {volume} {112}},\ \bibinfo {pages} {7449} (\bibinfo {year}
  {2015})}\BibitemShut {NoStop}%
\bibitem [{\citenamefont {Wang}\ \emph {et~al.}(2017)\citenamefont {Wang},
  \citenamefont {Guo}, \citenamefont {Xu}, \citenamefont {Zhang},\ and\
  \citenamefont {Chen}}]{Wang2017}%
  \BibitemOpen
  \bibfield  {author} {\bibinfo {author} {\bibfnamefont {F.}~\bibnamefont
  {Wang}}, \bibinfo {author} {\bibfnamefont {X.}~\bibnamefont {Guo}}, \bibinfo
  {author} {\bibfnamefont {J.}~\bibnamefont {Xu}}, \bibinfo {author}
  {\bibfnamefont {Y.}~\bibnamefont {Zhang}}, \ and\ \bibinfo {author}
  {\bibfnamefont {C.~Q.}\ \bibnamefont {Chen}},\ }\href {\doibase
  10.1115/1.4036476} {\bibfield  {journal} {\bibinfo  {journal} {J. Appl.
  Mech.}\ }\textbf {\bibinfo {volume} {84}},\ \bibinfo {pages} {061007}
  (\bibinfo {year} {2017})}\BibitemShut {NoStop}%
\bibitem [{\citenamefont {Cho}\ \emph {et~al.}(2014)\citenamefont {Cho},
  \citenamefont {Shin}, \citenamefont {Costa}, \citenamefont {Kim},
  \citenamefont {Kunin}, \citenamefont {Li}, \citenamefont {Lee}, \citenamefont
  {Yang}, \citenamefont {Han}, \citenamefont {Choi},\ and\ \citenamefont
  {Srolovitz}}]{Cho_PNAS_2014}%
  \BibitemOpen
  \bibfield  {author} {\bibinfo {author} {\bibfnamefont {Y.}~\bibnamefont
  {Cho}}, \bibinfo {author} {\bibfnamefont {J.-H.}\ \bibnamefont {Shin}},
  \bibinfo {author} {\bibfnamefont {A.}~\bibnamefont {Costa}}, \bibinfo
  {author} {\bibfnamefont {T.~A.}\ \bibnamefont {Kim}}, \bibinfo {author}
  {\bibfnamefont {V.}~\bibnamefont {Kunin}}, \bibinfo {author} {\bibfnamefont
  {J.}~\bibnamefont {Li}}, \bibinfo {author} {\bibfnamefont {S.~Y.}\
  \bibnamefont {Lee}}, \bibinfo {author} {\bibfnamefont {S.}~\bibnamefont
  {Yang}}, \bibinfo {author} {\bibfnamefont {H.~N.}\ \bibnamefont {Han}},
  \bibinfo {author} {\bibfnamefont {I.-S.}\ \bibnamefont {Choi}}, \ and\
  \bibinfo {author} {\bibfnamefont {D.~J.}\ \bibnamefont {Srolovitz}},\ }\href
  {\doibase 10.1073/pnas.1417276111} {\bibfield  {journal} {\bibinfo  {journal}
  {Proc. Natl. Acad. Sci. U.S.A.}\ }\textbf {\bibinfo {volume} {111}},\
  \bibinfo {pages} {17390} (\bibinfo {year} {2014})}\BibitemShut {NoStop}%
\bibitem [{\citenamefont {Rafsanjani}\ and\ \citenamefont
  {Pasini}(2016)}]{Rafsanjani2016}%
  \BibitemOpen
  \bibfield  {author} {\bibinfo {author} {\bibfnamefont {A.}~\bibnamefont
  {Rafsanjani}}\ and\ \bibinfo {author} {\bibfnamefont {D.}~\bibnamefont
  {Pasini}},\ }\href {\doibase 10.1016/j.eml.2016.09.001} {\bibfield  {journal}
  {\bibinfo  {journal} {Extreme Mech. Lett.}\ }\textbf {\bibinfo {volume}
  {9}},\ \bibinfo {pages} {291} (\bibinfo {year} {2016})}\BibitemShut {NoStop}%
\bibitem [{\citenamefont {Tang}\ and\ \citenamefont {Yin}(2017)}]{Tang2017}%
  \BibitemOpen
  \bibfield  {author} {\bibinfo {author} {\bibfnamefont {Y.}~\bibnamefont
  {Tang}}\ and\ \bibinfo {author} {\bibfnamefont {J.}~\bibnamefont {Yin}},\
  }\href {\doibase 10.1016/j.eml.2016.07.005} {\bibfield  {journal} {\bibinfo
  {journal} {Extreme Mech. Lett.}\ }\textbf {\bibinfo {volume} {12}},\ \bibinfo
  {pages} {77} (\bibinfo {year} {2017})}\BibitemShut {NoStop}%
\bibitem [{\citenamefont {Blees}\ \emph {et~al.}(2015)\citenamefont {Blees},
  \citenamefont {Barnard}, \citenamefont {Rose}, \citenamefont {Roberts},
  \citenamefont {McGill}, \citenamefont {Huang}, \citenamefont {Ruyack},
  \citenamefont {Kevek}, \citenamefont {Kobrin}, \citenamefont {Muller} \emph
  {et~al.}}]{Blees2015}%
  \BibitemOpen
  \bibfield  {author} {\bibinfo {author} {\bibfnamefont {M.~K.}\ \bibnamefont
  {Blees}}, \bibinfo {author} {\bibfnamefont {A.~W.}\ \bibnamefont {Barnard}},
  \bibinfo {author} {\bibfnamefont {P.~A.}\ \bibnamefont {Rose}}, \bibinfo
  {author} {\bibfnamefont {S.~P.}\ \bibnamefont {Roberts}}, \bibinfo {author}
  {\bibfnamefont {K.~L.}\ \bibnamefont {McGill}}, \bibinfo {author}
  {\bibfnamefont {P.~Y.}\ \bibnamefont {Huang}}, \bibinfo {author}
  {\bibfnamefont {A.~R.}\ \bibnamefont {Ruyack}}, \bibinfo {author}
  {\bibfnamefont {J.~W.}\ \bibnamefont {Kevek}}, \bibinfo {author}
  {\bibfnamefont {B.}~\bibnamefont {Kobrin}}, \bibinfo {author} {\bibfnamefont
  {D.~A.}\ \bibnamefont {Muller}},  \emph {et~al.},\ }\href {\doibase
  10.1038/nature14588} {\bibfield  {journal} {\bibinfo  {journal} {Nature}\
  }\textbf {\bibinfo {volume} {524}},\ \bibinfo {pages} {204} (\bibinfo {year}
  {2015})}\BibitemShut {NoStop}%
\bibitem [{\citenamefont {Rafsanjani}\ and\ \citenamefont
  {Bertoldi}(2017)}]{Rafsanjani2017}%
  \BibitemOpen
  \bibfield  {author} {\bibinfo {author} {\bibfnamefont {A.}~\bibnamefont
  {Rafsanjani}}\ and\ \bibinfo {author} {\bibfnamefont {K.}~\bibnamefont
  {Bertoldi}},\ }\href {\doibase 10.1103/PhysRevLett.118.084301} {\bibfield
  {journal} {\bibinfo  {journal} {Phys. Rev. Lett.}\ }\textbf {\bibinfo
  {volume} {118}},\ \bibinfo {pages} {084301} (\bibinfo {year}
  {2017})}\BibitemShut {NoStop}%
\bibitem [{\citenamefont {Dias}\ \emph {et~al.}(2017)\citenamefont {Dias},
  \citenamefont {McCarron}, \citenamefont {Rayneau-Kirkhope}, \citenamefont
  {Hanakata}, \citenamefont {Campbell}, \citenamefont {Park},\ and\
  \citenamefont {Holmes}}]{Dias_SOFTMATTER_2017}%
  \BibitemOpen
  \bibfield  {author} {\bibinfo {author} {\bibfnamefont {M.~A.}\ \bibnamefont
  {Dias}}, \bibinfo {author} {\bibfnamefont {M.~P.}\ \bibnamefont {McCarron}},
  \bibinfo {author} {\bibfnamefont {D.}~\bibnamefont {Rayneau-Kirkhope}},
  \bibinfo {author} {\bibfnamefont {P.~Z.}\ \bibnamefont {Hanakata}}, \bibinfo
  {author} {\bibfnamefont {D.~K.}\ \bibnamefont {Campbell}}, \bibinfo {author}
  {\bibfnamefont {H.~S.}\ \bibnamefont {Park}}, \ and\ \bibinfo {author}
  {\bibfnamefont {D.~P.}\ \bibnamefont {Holmes}},\ }\href {\doibase
  10.1039/C7SM01693J} {\bibfield  {journal} {\bibinfo  {journal} {Soft Matter}\
  }\textbf {\bibinfo {volume} {13}},\ \bibinfo {pages} {9087} (\bibinfo {year}
  {2017})}\BibitemShut {NoStop}%
\bibitem [{\citenamefont {Konakovi\'{c}-Lukovi\'{c}}\ \emph
  {et~al.}(2018)\citenamefont {Konakovi\'{c}-Lukovi\'{c}}, \citenamefont
  {Panetta}, \citenamefont {Crane},\ and\ \citenamefont
  {Pauly}}]{Konakovic2018}%
  \BibitemOpen
  \bibfield  {author} {\bibinfo {author} {\bibfnamefont {M.}~\bibnamefont
  {Konakovi\'{c}-Lukovi\'{c}}}, \bibinfo {author} {\bibfnamefont
  {J.}~\bibnamefont {Panetta}}, \bibinfo {author} {\bibfnamefont
  {K.}~\bibnamefont {Crane}}, \ and\ \bibinfo {author} {\bibfnamefont
  {M.}~\bibnamefont {Pauly}},\ }\href {\doibase 10.1145/3197517.3201373}
  {\bibfield  {journal} {\bibinfo  {journal} {ACM Trans. Graph.}\ }\textbf
  {\bibinfo {volume} {37}},\ \bibinfo {pages} {106} (\bibinfo {year}
  {2018})}\BibitemShut {NoStop}%
\bibitem [{\citenamefont {Celli}\ \emph {et~al.}(2018)\citenamefont {Celli},
  \citenamefont {McMahan}, \citenamefont {Ramirez}, \citenamefont {Bauhofer},
  \citenamefont {Naify}, \citenamefont {Hofmann}, \citenamefont {Audoly},\ and\
  \citenamefont {Daraio}}]{Celli2018}%
  \BibitemOpen
  \bibfield  {author} {\bibinfo {author} {\bibfnamefont {P.}~\bibnamefont
  {Celli}}, \bibinfo {author} {\bibfnamefont {C.}~\bibnamefont {McMahan}},
  \bibinfo {author} {\bibfnamefont {B.}~\bibnamefont {Ramirez}}, \bibinfo
  {author} {\bibfnamefont {A.}~\bibnamefont {Bauhofer}}, \bibinfo {author}
  {\bibfnamefont {C.}~\bibnamefont {Naify}}, \bibinfo {author} {\bibfnamefont
  {D.}~\bibnamefont {Hofmann}}, \bibinfo {author} {\bibfnamefont
  {B.}~\bibnamefont {Audoly}}, \ and\ \bibinfo {author} {\bibfnamefont
  {C.}~\bibnamefont {Daraio}},\ }\href {\doibase 10.1039/C8SM02082E} {\bibfield
   {journal} {\bibinfo  {journal} {Soft Matter}\ }\textbf {\bibinfo {volume}
  {14}},\ \bibinfo {pages} {9744} (\bibinfo {year} {2018})}\BibitemShut
  {NoStop}%
\bibitem [{\citenamefont {Choi}\ \emph {et~al.}(2019)\citenamefont {Choi},
  \citenamefont {Dudte},\ and\ \citenamefont {Mahadevan}}]{Choi2019}%
  \BibitemOpen
  \bibfield  {author} {\bibinfo {author} {\bibfnamefont {G.~P.~T.}\
  \bibnamefont {Choi}}, \bibinfo {author} {\bibfnamefont {L.~H.}\ \bibnamefont
  {Dudte}}, \ and\ \bibinfo {author} {\bibfnamefont {L.}~\bibnamefont
  {Mahadevan}},\ }\href {\doibase 10.1038/s41563-019-0452-y} {\bibfield
  {journal} {\bibinfo  {journal} {Nat. Mater.}\ }\textbf {\bibinfo {volume}
  {18}},\ \bibinfo {pages} {999} (\bibinfo {year} {2019})}\BibitemShut
  {NoStop}%
\bibitem [{\citenamefont {Yang}\ and\ \citenamefont
  {You}(2018)}]{Yang2018geometry}%
  \BibitemOpen
  \bibfield  {author} {\bibinfo {author} {\bibfnamefont {Y.}~\bibnamefont
  {Yang}}\ and\ \bibinfo {author} {\bibfnamefont {Z.}~\bibnamefont {You}},\
  }\href {\doibase 10.1115/1.4038969} {\bibfield  {journal} {\bibinfo
  {journal} {J. Mech. Robot.}\ }\textbf {\bibinfo {volume} {10}},\ \bibinfo
  {pages} {021001} (\bibinfo {year} {2018})}\BibitemShut {NoStop}%
\bibitem [{\citenamefont {Singh}\ and\ \citenamefont {van
  Hecke}(2021)}]{Singh2021}%
  \BibitemOpen
  \bibfield  {author} {\bibinfo {author} {\bibfnamefont {N.}~\bibnamefont
  {Singh}}\ and\ \bibinfo {author} {\bibfnamefont {M.}~\bibnamefont {van
  Hecke}},\ }\href {\doibase 10.1103/PhysRevLett.126.248002} {\bibfield
  {journal} {\bibinfo  {journal} {Phys. Rev. Lett.}\ }\textbf {\bibinfo
  {volume} {126}},\ \bibinfo {pages} {248002} (\bibinfo {year}
  {2021})}\BibitemShut {NoStop}%
\bibitem [{\citenamefont {Dang}\ \emph
  {et~al.}(2022{\natexlab{a}})\citenamefont {Dang}, \citenamefont {Feng},
  \citenamefont {Duan},\ and\ \citenamefont {Wang}}]{dang2022theorem}%
  \BibitemOpen
  \bibfield  {author} {\bibinfo {author} {\bibfnamefont {X.}~\bibnamefont
  {Dang}}, \bibinfo {author} {\bibfnamefont {F.}~\bibnamefont {Feng}}, \bibinfo
  {author} {\bibfnamefont {H.}~\bibnamefont {Duan}}, \ and\ \bibinfo {author}
  {\bibfnamefont {J.}~\bibnamefont {Wang}},\ }\href {\doibase
  10.1103/PhysRevLett.128.035501} {\bibfield  {journal} {\bibinfo  {journal}
  {Phys. Rev. Lett.}\ }\textbf {\bibinfo {volume} {128}},\ \bibinfo {pages}
  {035501} (\bibinfo {year} {2022}{\natexlab{a}})}\BibitemShut {NoStop}%
\bibitem [{\citenamefont {Courant}\ and\ \citenamefont
  {Hilbert}(2008)}]{courant2008methods}%
  \BibitemOpen
  \bibfield  {author} {\bibinfo {author} {\bibfnamefont {R.}~\bibnamefont
  {Courant}}\ and\ \bibinfo {author} {\bibfnamefont {D.}~\bibnamefont
  {Hilbert}},\ }\href@noop {} {\emph {\bibinfo {title} {Methods of mathematical
  physics: partial differential equations}}}\ (\bibinfo  {publisher} {John
  Wiley \& Sons},\ \bibinfo {year} {2008})\BibitemShut {NoStop}%
\bibitem [{\citenamefont {Evans}(2010)}]{evans10}%
  \BibitemOpen
  \bibfield  {author} {\bibinfo {author} {\bibfnamefont {L.~C.}\ \bibnamefont
  {Evans}},\ }\href@noop {} {\emph {\bibinfo {title} {Partial differential
  equations}}}\ (\bibinfo  {publisher} {American Mathematical Society},\
  \bibinfo {address} {Providence, R.I.},\ \bibinfo {year} {2010})\BibitemShut
  {NoStop}%
\bibitem [{\citenamefont {Nassar}\ \emph {et~al.}(2017)\citenamefont {Nassar},
  \citenamefont {Leb{\'e}e},\ and\ \citenamefont
  {Monasse}}]{nassar2017curvature}%
  \BibitemOpen
  \bibfield  {author} {\bibinfo {author} {\bibfnamefont {H.}~\bibnamefont
  {Nassar}}, \bibinfo {author} {\bibfnamefont {A.}~\bibnamefont {Leb{\'e}e}}, \
  and\ \bibinfo {author} {\bibfnamefont {L.}~\bibnamefont {Monasse}},\
  }\href@noop {} {\bibfield  {journal} {\bibinfo  {journal} {Proc. Royal Soc.
  A}\ }\textbf {\bibinfo {volume} {473}},\ \bibinfo {pages} {20160705}
  (\bibinfo {year} {2017})}\BibitemShut {NoStop}%
\bibitem [{\citenamefont {Leb{\'e}e}\ \emph {et~al.}(2018)\citenamefont
  {Leb{\'e}e}, \citenamefont {Monasse},\ and\ \citenamefont
  {Nassar}}]{lebee2018fitting}%
  \BibitemOpen
  \bibfield  {author} {\bibinfo {author} {\bibfnamefont {A.}~\bibnamefont
  {Leb{\'e}e}}, \bibinfo {author} {\bibfnamefont {L.}~\bibnamefont {Monasse}},
  \ and\ \bibinfo {author} {\bibfnamefont {H.}~\bibnamefont {Nassar}},\ }in\
  \href@noop {} {\emph {\bibinfo {booktitle} {7th International Meeting on
  Origami in Science, Mathematics and Education (7OSME)}}},\ Vol.~\bibinfo
  {volume} {4}\ (\bibinfo {organization} {Tarquin},\ \bibinfo {year} {2018})\
  p.\ \bibinfo {pages} {811}\BibitemShut {NoStop}%
\bibitem [{\citenamefont {Rocklin}\ \emph {et~al.}(2017)\citenamefont
  {Rocklin}, \citenamefont {Zhou}, \citenamefont {Sun},\ and\ \citenamefont
  {Mao}}]{rocklin2017transformable}%
  \BibitemOpen
  \bibfield  {author} {\bibinfo {author} {\bibfnamefont {D.~Z.}\ \bibnamefont
  {Rocklin}}, \bibinfo {author} {\bibfnamefont {S.}~\bibnamefont {Zhou}},
  \bibinfo {author} {\bibfnamefont {K.}~\bibnamefont {Sun}}, \ and\ \bibinfo
  {author} {\bibfnamefont {X.}~\bibnamefont {Mao}},\ }\href@noop {} {\bibfield
  {journal} {\bibinfo  {journal} {Nat. Commun.}\ }\textbf {\bibinfo {volume}
  {8}},\ \bibinfo {pages} {1} (\bibinfo {year} {2017})}\BibitemShut {NoStop}%
\bibitem [{sup()}]{suppl}%
  \BibitemOpen
  \href@noop {} {}\bibinfo {note} {See Supplemental Material for further
  theoretical and experimental details.}\BibitemShut {Stop}%
\bibitem [{\citenamefont {Courant}\ and\ \citenamefont
  {Friedrichs}(1999)}]{courant1999supersonic}%
  \BibitemOpen
  \bibfield  {author} {\bibinfo {author} {\bibfnamefont {R.}~\bibnamefont
  {Courant}}\ and\ \bibinfo {author} {\bibfnamefont {K.~O.}\ \bibnamefont
  {Friedrichs}},\ }\href@noop {} {\emph {\bibinfo {title} {Supersonic flow and
  shock waves}}},\ Vol.~\bibinfo {volume} {21}\ (\bibinfo  {publisher}
  {Springer Science \& Business Media},\ \bibinfo {year} {1999})\BibitemShut
  {NoStop}%
\bibitem [{\citenamefont {Do~Carmo}(2016)}]{do2016differential}%
  \BibitemOpen
  \bibfield  {author} {\bibinfo {author} {\bibfnamefont {M.~P.}\ \bibnamefont
  {Do~Carmo}},\ }\href@noop {} {\emph {\bibinfo {title} {Differential geometry
  of curves and surfaces: revised and updated second edition}}}\ (\bibinfo
  {publisher} {Courier Dover Publications},\ \bibinfo {year}
  {2016})\BibitemShut {NoStop}%
\bibitem [{\citenamefont {Brown}\ and\ \citenamefont
  {Churchill}(2009)}]{brown2009complex}%
  \BibitemOpen
  \bibfield  {author} {\bibinfo {author} {\bibfnamefont {J.~W.}\ \bibnamefont
  {Brown}}\ and\ \bibinfo {author} {\bibfnamefont {R.~V.}\ \bibnamefont
  {Churchill}},\ }\href@noop {} {\emph {\bibinfo {title} {Complex variables and
  applications eighth edition}}}\ (\bibinfo  {publisher} {McGraw-Hill Book
  Company},\ \bibinfo {year} {2009})\BibitemShut {NoStop}%
\bibitem [{\citenamefont {Dudte}\ \emph {et~al.}(2016)\citenamefont {Dudte},
  \citenamefont {Vouga}, \citenamefont {Tachi},\ and\ \citenamefont
  {Mahadevan}}]{Dudte2016}%
  \BibitemOpen
  \bibfield  {author} {\bibinfo {author} {\bibfnamefont {L.~H.}\ \bibnamefont
  {Dudte}}, \bibinfo {author} {\bibfnamefont {E.}~\bibnamefont {Vouga}},
  \bibinfo {author} {\bibfnamefont {T.}~\bibnamefont {Tachi}}, \ and\ \bibinfo
  {author} {\bibfnamefont {L.}~\bibnamefont {Mahadevan}},\ }\href {\doibase
  10.1038/NMAT4540} {\bibfield  {journal} {\bibinfo  {journal} {Nat. Mater.}\
  }\textbf {\bibinfo {volume} {15}},\ \bibinfo {pages} {583} (\bibinfo {year}
  {2016})}\BibitemShut {NoStop}%
\bibitem [{\citenamefont {Dang}\ \emph
  {et~al.}(2022{\natexlab{b}})\citenamefont {Dang}, \citenamefont {Feng},
  \citenamefont {Plucinsky}, \citenamefont {James}, \citenamefont {Duan},\ and\
  \citenamefont {Wang}}]{dang2022inverse}%
  \BibitemOpen
  \bibfield  {author} {\bibinfo {author} {\bibfnamefont {X.}~\bibnamefont
  {Dang}}, \bibinfo {author} {\bibfnamefont {F.}~\bibnamefont {Feng}}, \bibinfo
  {author} {\bibfnamefont {P.}~\bibnamefont {Plucinsky}}, \bibinfo {author}
  {\bibfnamefont {R.~D.}\ \bibnamefont {James}}, \bibinfo {author}
  {\bibfnamefont {H.}~\bibnamefont {Duan}}, \ and\ \bibinfo {author}
  {\bibfnamefont {J.}~\bibnamefont {Wang}},\ }\href@noop {} {\bibfield
  {journal} {\bibinfo  {journal} {International Journal of Solids and
  Structures}\ }\textbf {\bibinfo {volume} {234}},\ \bibinfo {pages} {111224}
  (\bibinfo {year} {2022}{\natexlab{b}})}\BibitemShut {NoStop}%
\bibitem [{\citenamefont {Khajehtourian}\ and\ \citenamefont
  {Kochmann}(2021)}]{khajehtourian2021continuum}%
  \BibitemOpen
  \bibfield  {author} {\bibinfo {author} {\bibfnamefont {R.}~\bibnamefont
  {Khajehtourian}}\ and\ \bibinfo {author} {\bibfnamefont {D.~M.}\ \bibnamefont
  {Kochmann}},\ }\href {\doibase 10.1016/j.jmps.2020.104217} {\bibfield
  {journal} {\bibinfo  {journal} {J. Mech. Phys. Solids}\ }\textbf {\bibinfo
  {volume} {147}},\ \bibinfo {pages} {104217} (\bibinfo {year}
  {2021})}\BibitemShut {NoStop}%
\bibitem [{\citenamefont {McMahan}\ \emph {et~al.}(2021)\citenamefont
  {McMahan}, \citenamefont {Akerson}, \citenamefont {Celli}, \citenamefont
  {Audoly},\ and\ \citenamefont {Daraio}}]{mcmahan2021effective}%
  \BibitemOpen
  \bibfield  {author} {\bibinfo {author} {\bibfnamefont {C.}~\bibnamefont
  {McMahan}}, \bibinfo {author} {\bibfnamefont {A.}~\bibnamefont {Akerson}},
  \bibinfo {author} {\bibfnamefont {P.}~\bibnamefont {Celli}}, \bibinfo
  {author} {\bibfnamefont {B.}~\bibnamefont {Audoly}}, \ and\ \bibinfo {author}
  {\bibfnamefont {C.}~\bibnamefont {Daraio}},\ }\href@noop {} {\bibfield
  {journal} {\bibinfo  {journal} {arXiv preprint arXiv:2107.01704}\ } (\bibinfo
  {year} {2021})}\BibitemShut {NoStop}%
\bibitem [{\citenamefont {Deng}\ \emph {et~al.}(2017)\citenamefont {Deng},
  \citenamefont {Raney}, \citenamefont {Tournat},\ and\ \citenamefont
  {Bertoldi}}]{Deng2017}%
  \BibitemOpen
  \bibfield  {author} {\bibinfo {author} {\bibfnamefont {B.}~\bibnamefont
  {Deng}}, \bibinfo {author} {\bibfnamefont {J.~R.}\ \bibnamefont {Raney}},
  \bibinfo {author} {\bibfnamefont {V.}~\bibnamefont {Tournat}}, \ and\
  \bibinfo {author} {\bibfnamefont {K.}~\bibnamefont {Bertoldi}},\ }\href
  {\doibase 10.1103/PhysRevLett.118.204102} {\bibfield  {journal} {\bibinfo
  {journal} {Phys. Rev. Lett.}\ }\textbf {\bibinfo {volume} {118}},\ \bibinfo
  {pages} {204102} (\bibinfo {year} {2017})}\BibitemShut {NoStop}%
\end{thebibliography}%


\begin{thebibliography}{5}%
\makeatletter
\providecommand \@ifxundefined [1]{%
 \@ifx{#1\undefined}
}%
\providecommand \@ifnum [1]{%
 \ifnum #1\expandafter \@firstoftwo
 \else \expandafter \@secondoftwo
 \fi
}%
\providecommand \@ifx [1]{%
 \ifx #1\expandafter \@firstoftwo
 \else \expandafter \@secondoftwo
 \fi
}%
\providecommand \natexlab [1]{#1}%
\providecommand \enquote  [1]{``#1''}%
\providecommand \bibnamefont  [1]{#1}%
\providecommand \bibfnamefont [1]{#1}%
\providecommand \citenamefont [1]{#1}%
\providecommand \href@noop [0]{\@secondoftwo}%
\providecommand \href [0]{\begingroup \@sanitize@url \@href}%
\providecommand \@href[1]{\@@startlink{#1}\@@href}%
\providecommand \@@href[1]{\endgroup#1\@@endlink}%
\providecommand \@sanitize@url [0]{\catcode `\\12\catcode `\$12\catcode
  `\&12\catcode `\#12\catcode `\^12\catcode `\_12\catcode `\%12\relax}%
\providecommand \@@startlink[1]{}%
\providecommand \@@endlink[0]{}%
\providecommand \url  [0]{\begingroup\@sanitize@url \@url }%
\providecommand \@url [1]{\endgroup\@href {#1}{\urlprefix }}%
\providecommand \urlprefix  [0]{URL }%
\providecommand \Eprint [0]{\href }%
\providecommand \doibase [0]{http://dx.doi.org/}%
\providecommand \selectlanguage [0]{\@gobble}%
\providecommand \bibinfo  [0]{\@secondoftwo}%
\providecommand \bibfield  [0]{\@secondoftwo}%
\providecommand \translation [1]{[#1]}%
\providecommand \BibitemOpen [0]{}%
\providecommand \bibitemStop [0]{}%
\providecommand \bibitemNoStop [0]{.\EOS\space}%
\providecommand \EOS [0]{\spacefactor3000\relax}%
\providecommand \BibitemShut  [1]{\csname bibitem#1\endcsname}%
\let\auto@bib@innerbib\@empty
\bibitem [{\citenamefont {Deng}\ \emph {et~al.}(2020)\citenamefont {Deng},
  \citenamefont {Yu}, \citenamefont {Forte}, \citenamefont {Tournat},\ and\
  \citenamefont {Bertoldi}}]{Deng2020}%
  \BibitemOpen
  \bibfield  {author} {\bibinfo {author} {\bibfnamefont {B.}~\bibnamefont
  {Deng}}, \bibinfo {author} {\bibfnamefont {S.}~\bibnamefont {Yu}}, \bibinfo
  {author} {\bibfnamefont {A.~E.}\ \bibnamefont {Forte}}, \bibinfo {author}
  {\bibfnamefont {V.}~\bibnamefont {Tournat}}, \ and\ \bibinfo {author}
  {\bibfnamefont {K.}~\bibnamefont {Bertoldi}},\ }\href {\doibase
  10.1073/pnas.2015847117} {\bibfield  {journal} {\bibinfo  {journal} {Proc.
  Natl. Acad. Sci. U.S.A.}\ }\textbf {\bibinfo {volume} {117}},\ \bibinfo
  {pages} {31002} (\bibinfo {year} {2020})}\BibitemShut {NoStop}%
\bibitem [{\citenamefont {Czajkowski}\ \emph {et~al.}(2022)\citenamefont
  {Czajkowski}, \citenamefont {Coulais}, \citenamefont {van Hecke},\ and\
  \citenamefont {Rocklin}}]{czajkowski2022conformal}%
  \BibitemOpen
  \bibfield  {author} {\bibinfo {author} {\bibfnamefont {M.}~\bibnamefont
  {Czajkowski}}, \bibinfo {author} {\bibfnamefont {C.}~\bibnamefont {Coulais}},
  \bibinfo {author} {\bibfnamefont {M.}~\bibnamefont {van Hecke}}, \ and\
  \bibinfo {author} {\bibfnamefont {D.~Z.}\ \bibnamefont {Rocklin}},\ }\href
  {\doibase 10.1038/s41467-021-27825-0} {\bibfield  {journal} {\bibinfo
  {journal} {Nat. Commun.}\ }\textbf {\bibinfo {volume} {13}},\ \bibinfo
  {pages} {211} (\bibinfo {year} {2022})}\BibitemShut {NoStop}%
\bibitem [{\citenamefont {Courant}\ and\ \citenamefont
  {Hilbert}(2008)}]{courant2008methods}%
  \BibitemOpen
  \bibfield  {author} {\bibinfo {author} {\bibfnamefont {R.}~\bibnamefont
  {Courant}}\ and\ \bibinfo {author} {\bibfnamefont {D.}~\bibnamefont
  {Hilbert}},\ }\href@noop {} {\emph {\bibinfo {title} {Methods of mathematical
  physics: partial differential equations}}}\ (\bibinfo  {publisher} {John
  Wiley \& Sons},\ \bibinfo {year} {2008})\BibitemShut {NoStop}%
\bibitem [{\citenamefont {Lax}(2007)}]{lax07}%
  \BibitemOpen
  \bibfield  {author} {\bibinfo {author} {\bibfnamefont {P.~D.}\ \bibnamefont
  {Lax}},\ }\href@noop {} {\emph {\bibinfo {title} {Linear Algebra and Its
  Applications}}},\ \bibinfo {edition} {2nd}\ ed.\ (\bibinfo  {publisher}
  {Wiley-Interscience},\ \bibinfo {address} {Hoboken, NJ},\ \bibinfo {year}
  {2007})\BibitemShut {NoStop}%
\bibitem [{\citenamefont {Evans}(2010)}]{evans10}%
  \BibitemOpen
  \bibfield  {author} {\bibinfo {author} {\bibfnamefont {L.~C.}\ \bibnamefont
  {Evans}},\ }\href@noop {} {\emph {\bibinfo {title} {Partial differential
  equations}}}\ (\bibinfo  {publisher} {American Mathematical Society},\
  \bibinfo {address} {Providence, R.I.},\ \bibinfo {year} {2010})\BibitemShut
  {NoStop}%
\end{thebibliography}%

\end{document}


\begin{center}
\Large
{\textbf{Supplemental Material (SM): Continuum field theory for the deformations of planar kirigami}}
\vspace{2mm}
\large

\vspace{2mm}
Yue Zheng, Imtiar Niloy, Paolo Celli,  Ian Tobasco and Paul Plucinsky
\end{center}

\renewcommand{\thesection}{SM.\arabic{section}} 
\renewcommand{\thefigure}{S\arabic{figure}} 
\renewcommand{\thepage}{S\arabic{page}}  
\renewcommand{\theequation}{S\arabic{equation}}  

\tableofcontents

\vspace{.5cm}
\noindent \textbf{Notation.}\;We refer to equations and figures from the main text as Eq.\;(1), Fig.\;1 and so on. The SM versions are distinguished by an ``S", as in Eq.\;(S1), Fig.\;S1.  

\section{Unit cells of four quad panels and four parallelogram slits}
\label{s:UC}

\begin{figure}[h!]
\centering
\includegraphics[width =\linewidth]{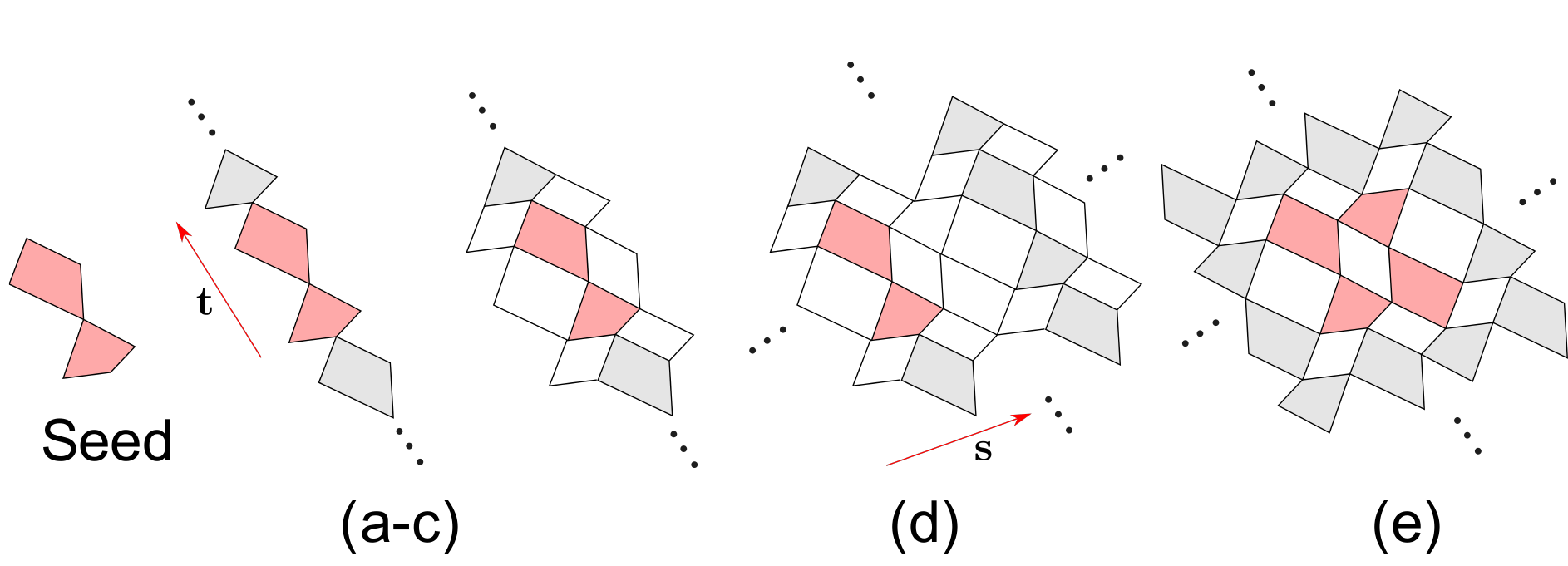}
\caption{Geometric argument for constructing unit cells with quad panels and parallelogram slits. }
\label{fig:GeomArg}
\end{figure}

In this section, we derive all possible unit cells in quad-kirigami composed of four convex quad panels and four parallelogram slits.  The result is based on the geometric argument in Fig.\;\ref{fig:GeomArg}, which establishes a series of necessary conditions that eventually leads to the overall characterization. 

We start with two generic quad panels connected at a corner (Fig.\;\ref{fig:GeomArg}(a)). We will show that these two panels \textit{seed} the entire pattern.  First, observe that the two panels must repeat along the vector indicated by $\mathbf{t}$ (Fig.\;\ref{fig:GeomArg}(b)), since our unit cell consists of four quad panels. As the slits are assumed to be parallelograms, we can conclude their shapes from two of their sides (Fig.\;\ref{fig:GeomArg}(c)).  We now have identified the four parallelogram slits of our unit cell. The rest of the pattern is obtained through a tessellation (Fig.\;\ref{fig:GeomArg}(d)), in this case along the vector $\mathbf{s}$. This identifies the final two panels of the unit cell (Fig.\;\ref{fig:GeomArg}(e)), which turn out to be a rotation of the seed by $180^{\circ}$.  

As a word of caution, we note that it is possible for a given seed to initiate an invalid tessellation, in the sense that two of the resulting panels overlap. One can avoid this by counter-rotating the seeded panels until a valid configuration is obtained. 

\section{Mechanism deformations}\label{sec:KinAppend}

Here we  describe all possible planar and \textit{mechanism deformations} of our kirigami, i.e., those that  rotate and translate the panels in the plane while preserving the pattern's topology. Coarse-graining leads to the effective description in Eq.\;(3-4) of the main text. We specialize to kirigami with rhombi-slits at the end.

\subsection{Identifying mechanisms}
\begin{figure}[t]
\centering
\includegraphics[width = 4in]{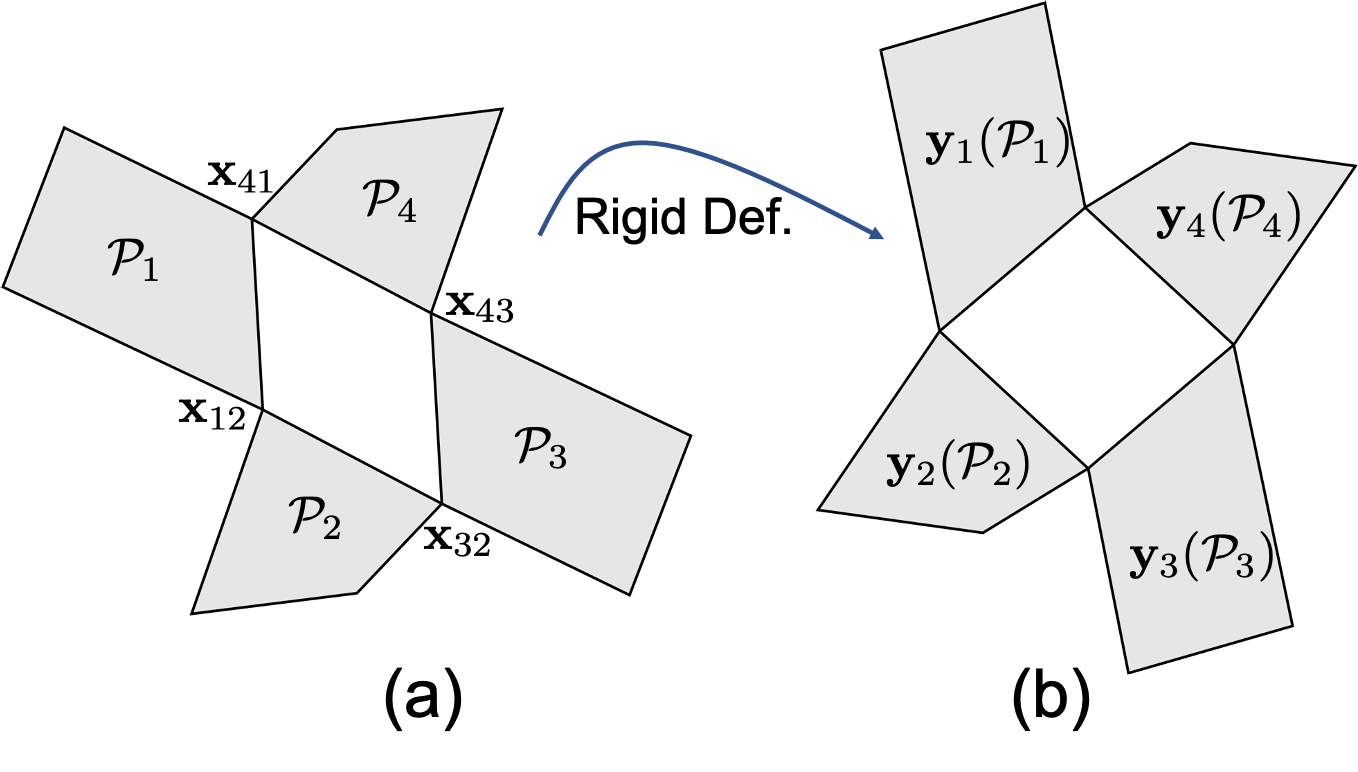}
\caption{Notation for rigidly deforming a unit cell of the kirigami.}
\label{fig:DescribeKin}
\end{figure}

We first study a single unit cell of four panels $\mathcal{P}_i$, labeled as shown in Fig.\;\ref{fig:DescribeKin}, with each $\mathbf{x}_{ij}$ indicating the corner points connecting adjacent  panels. 
The rigid deformations of this cell have the form 
\begin{equation}
\begin{aligned}\label{eq:Rigid0}
\mathbf{y}_i(\mathbf{x}) = \mathbf{R}_{i} \mathbf{x}+ \mathbf{c}_i, \quad \mathbf{x} \in \mathcal{P}_i, \quad i = 1,\ldots,4
\end{aligned}
\end{equation}
for appropriately chosen 2D rotations $\mathbf{R}_i$ and 2D translations $\mathbf{c}_i$.  In particular, to preserve the topology, the deformations must satisfy 
\begin{equation}
\begin{aligned}\label{eq:connect}
\mathbf{y}_{1}(\mathbf{x}_{41}) = \mathbf{y}_4(\mathbf{x}_{41}), \quad \mathbf{y}_2(\mathbf{x}_{12}) =\mathbf{y}_1(\mathbf{x}_{12}), \quad \mathbf{y}_{3}(\mathbf{x}_{32}) = \mathbf{y}_{2}(\mathbf{x}_{32}), \quad \mathbf{y}_{4}(\mathbf{x}_{43}) = \mathbf{y}_{3}(\mathbf{x}_{43}).
\end{aligned}
\end{equation}
By routine manipulation, this restriction is equivalent to prescribing three of the four translations as 
\begin{equation}
\begin{aligned}\label{eq:c123}
&\mathbf{c}_1 = (\mathbf{R}_4 - \mathbf{R}_1) \mathbf{x}_{41} + \mathbf{c}_4, \\
&\mathbf{c}_2 = (\mathbf{R}_4 - \mathbf{R}_1) \mathbf{x}_{41} + (\mathbf{R}_1 - \mathbf{R}_2) \mathbf{x}_{12} +  \mathbf{c}_4, \\
&\mathbf{c}_3 = (\mathbf{R}_4 - \mathbf{R}_1) \mathbf{x}_{41} + (\mathbf{R}_1 - \mathbf{R}_2)  \mathbf{x}_{12} + (\mathbf{R}_2 - \mathbf{R}_3) \mathbf{x}_{32} + \mathbf{c}_4
\end{aligned}
\end{equation}
and solving the \textit{loop compatibility condition}
\begin{equation}
\begin{aligned}\label{eq:connectFinal}
\mathbf{R}_1( \mathbf{x}_{12} - \mathbf{x}_{41}) + \mathbf{R}_{2}(\mathbf{x}_{32} - \mathbf{x}_{12}) + \mathbf{R}_3(\mathbf{x}_{43} - \mathbf{x}_{32}) + \mathbf{R}_4 (\mathbf{x}_{41} - \mathbf{x}_{43}) = \mathbf{0}.
\end{aligned}
\end{equation}
Concerning the nature of  (\ref{eq:connectFinal}), we note that each term in the sum is a vector describing one side of the deformed slit.  Summing the terms, as shown, traverses the sides in a loop that must return back to the starting point for the slit to remain closed.
The analysis, so far, is completely general and applies regardless of the shape of the slits.

The next step is to solve for the rotations $\mathbf{R}_i, i = 1,\ldots, 4$, and build on this solution to construct the rigid panel deformations. Here, as it is consistent with the patterns in the main text, we simplify to the case of a parallelogram slit given by the vectors
\begin{equation}
\begin{aligned}\label{eq:parallelProp}
\mathbf{u} = \mathbf{x}_{12} - \mathbf{x}_{41} = -(\mathbf{x}_{43} - \mathbf{x}_{32}) , \quad \mathbf{v} = \mathbf{x}_{32} - \mathbf{x}_{12} = -( \mathbf{x}_{41} - \mathbf{x}_{43}).
\end{aligned}
\end{equation}
The loop condition reduces to 
\begin{equation}
\begin{aligned}\label{eq:loopParallel}
(\mathbf{R}_1 - \mathbf{R}_3) \mathbf{u} + (\mathbf{R}_2 - \mathbf{R}_4) \mathbf{v} = \mathbf{0}
\end{aligned}
\end{equation}
and we read off the solution $\mathbf{R}_1 = \mathbf{R}_3$ and $\mathbf{R}_2 = \mathbf{R}_4$. That this is the only solution for the motion of the parallelogram slit follows from elementary geometry (e.g., by a routine application of the law of cosines). 
Thus, the rigid deformations of the unit cell are parameterized by rotations that satisfy
\begin{equation}\label{eq:rotations-result}
\begin{aligned}
\mathbf{R}_1 = \mathbf{R}_3 = \mathbf{R}(\gamma + \xi), \quad \mathbf{R}_2 = \mathbf{R}_4 = \mathbf{R}(\gamma - \xi)
\end{aligned}
\end{equation}
for two angles $\gamma$ and $\xi$, and translations $\mathbf{c}_{i}$, $i=1,2,3$ obeying (\ref{eq:c123}).  The angle $\gamma$ rotates the unit cell as a whole. The angle $\xi$ describes how its slit opens and closes. The slit angle $\angle \mathbf{x}_{41}\mathbf{x}_{12} \mathbf{x}_{32}$ in Fig.\;\ref{fig:DescribeKin} is deformed to $\angle \mathbf{x}_{41}\mathbf{x}_{12} \mathbf{x}_{32} + 2 \xi$ under our parameterization. 


\begin{figure}[b]
\centering
\includegraphics[width = 7in]{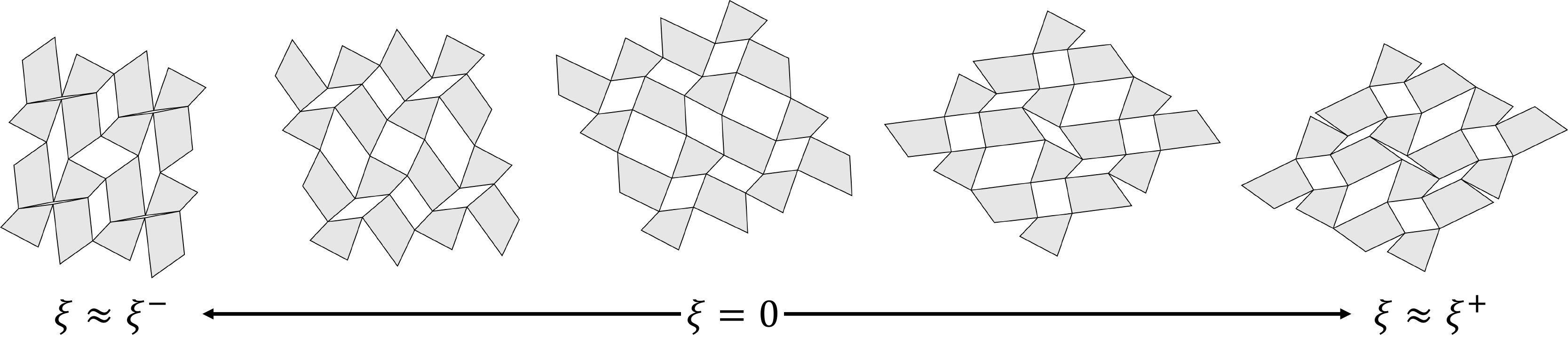}
\caption{Kirigami patterns with parallelogram slits  can deform as a mechanism in the plane.}
\label{fig:MechDef}
\end{figure}

Iterating the result of (\ref{eq:rotations-result}) yields a characterization of all rigid mechanism deformations of the kirigami: its panel rotations must alternate between $\mathbf{R}(\gamma + \xi)$ and $\mathbf{R}(\gamma -\xi)$ throughout. For a given set of alternating rotations, the corresponding panel translations that preserve the pattern's topology are uniquely prescribed up to an overall translation. Since $\xi$ is an angle describing how the slits open and close, it has a maximum possible value $\xi^{+} \in [0, \pi/2]$ and a minimum possible value $\xi^{-} \in [-\pi/2, 0]$ beyond which continued deformation results in  panels overlapping. In a valid mechanism, $\xi \in [\xi^{-}, \xi^+]$. Fig.\;\ref{fig:MechDef} depicts this result for the pattern in Fig.\;\ref{fig:GeomArg}(e).  The deformation continuously evolves as a function of $\xi$, distorting the slits while keeping the panels rigid until eventually one set of slits closes (at $\xi=\xi^-$ or $\xi=\xi^+$). 

\subsection{Coarse-grained description of mechanisms} In the main text, we coarse-grained the mechanism deformation discussed above to derive its \textit{effective deformation gradient} 
\begin{equation}
    \begin{aligned}
    \mathbf{F}_{\text{eff}} = \mathbf{R}(\gamma) \mathbf{A}(\xi).
    \end{aligned}
\end{equation}
To reiterate, this formula captures the bulk behavior of the mechanism  parameterized by the angles $\gamma$ and $\xi$. It applies to any kirigami pattern composed of a unit cell of four quad panels and four parallelograms slits. Recall the vectors $\mathbf{s}_i$, $\mathbf{t}_i$, $i = 1,\ldots,4$, from Fig.\;2.  From these vectors and Eq.\;(1), we find that 
\begin{equation}
    \begin{aligned}\label{eq:AxiExplicit}
    \mathbf{A}(\xi) = \mathbf{a}(\xi) \otimes \mathbf{s} + \mathbf{b}(\xi) \otimes \mathbf{W} \mathbf{s}
    \end{aligned}
\end{equation}
where $\mathbf{W} = \mathbf{R}(\pi/2)$, and 
\begin{equation}
\begin{aligned}\label{eq:abKin}
&\mathbf{a}(\xi) = \mathbf{R}(-\xi) \mathbf{a}_1 + \mathbf{R}(\xi) \mathbf{a}_2, \quad \mathbf{b}(\xi) = \mathbf{R}(-\xi) \mathbf{b}_1 + \mathbf{R}(\xi) \mathbf{b}_2.
\end{aligned}
\end{equation}
The vectors $\mathbf{a}_{1,2}$ and $\mathbf{b}_{1,2}$ encode the shape of the reference pattern explicitly via
\begin{equation}
\begin{aligned}\label{eq:abShape}
&\mathbf{a}_1 = |\mathbf{s}|^{-2} \mathbf{s}_{12}, && \mathbf{b}_1 =   \tfrac{1}{ (\mathbf{W} \mathbf{s} \cdot \mathbf{t})} \mathbf{t}_{14}    - \tfrac{(\mathbf{s} \cdot \mathbf{t})}{|\mathbf{s}|^2 (\mathbf{W} \mathbf{s} \cdot \mathbf{t})} \mathbf{s}_{12},  \\
&\mathbf{a}_2 = |\mathbf{s}|^{-2} \mathbf{s}_{34}, 
&& \mathbf{b}_2 =   \tfrac{1}{ (\mathbf{W} \mathbf{s} \cdot \mathbf{t})} \mathbf{t}_{23}    - \tfrac{(\mathbf{s} \cdot \mathbf{t})}{|\mathbf{s}|^2 (\mathbf{W} \mathbf{s} \cdot \mathbf{t})} \mathbf{s}_{34}
\end{aligned}
\end{equation} 
where $\mathbf{s}_{ij} = \mathbf{s}_i + \mathbf{s}_j$ and $\mathbf{t}_{ij} = \mathbf{t}_i + \mathbf{t}_j$.

\subsection{Simplification to rhombi-slits} The examples in the main text focus on the special case of kirigami with rhombi-slits. Such patterns correspond to certain special choices of $\mathbf{s}_i,\mathbf{t}_i$:
\begin{equation}
    \begin{aligned}\label{eq:sVtVExplicit}
&2\mathbf{s}_1 = (1-\lambda_3) \mathbf{e}_1 -  a_r (1-\lambda_2) \mathbf{e}_2 , && 2\mathbf{s}_4 =(1-\lambda_3) \mathbf{e}_1 +  a_r (1-\lambda_2) \mathbf{e}_2 , \\
&2\mathbf{s}_2 =  \lambda_3 \mathbf{e}_1 + a_r \lambda_2 \mathbf{e}_2,  &&  2\mathbf{s}_3 =  \lambda_3 \mathbf{e}_1 - a_r \lambda_2 \mathbf{e}_2 , \\
&2\mathbf{t}_1 =  a_r \lambda_2 \mathbf{e}_2 + (1- \lambda_1) \mathbf{e}_1,    && 2\mathbf{t}_2 =  a_r \lambda_2 \mathbf{e}_2 - (1- \lambda_1) \mathbf{e}_1, \\
&2 \mathbf{t}_3 = a_r (1-\lambda_2) \mathbf{e}_2 + (1-\lambda_3) \mathbf{e}_1,   && 2\mathbf{t}_4 = a_r (1-\lambda_2) \mathbf{e}_2 - (1-\lambda_3) \mathbf{e}_1, 
    \end{aligned}
\end{equation}
for $\lambda_1,\ldots, \lambda_4$ taking values in $[0,1]$, and an aspect ratio $a_r >0$. We use $\mathbf{e}_1, \mathbf{e}_2$ for the standard Cartesian basis vectors on $\mathbb{R}^2$. These formulas follow by comparing the vectors $\mathbf{s}_1,\ldots, \mathbf{t}_4$ defining the unit cell in Fig.\;2 to the $(\lambda_1, \ldots, \lambda_4,a_r)$-parameterization of the rhombi-slit cell in Fig.\;3. Observe that $\mathbf{s} = \sum_{i=1,\ldots,4} \mathbf{s}_i = \mathbf{e}_1$ and $\mathbf{t} =\sum_{i=1,\ldots,4} \mathbf{t}_i = a_r \mathbf{e}_2$. The explicit formula for $\mathbf{A}(\xi)$ in Eq.\;(5) is obtained by substituting  (\ref{eq:sVtVExplicit}) into (\ref{eq:AxiExplicit}-\ref{eq:abShape}) and performing routine algebra. 

In fact, the formulas in (\ref{eq:sVtVExplicit}) describe \textit{all} kirigami patterns with a unit cell of four quad panels and four rhombi-slits. This follows from the argument in SM.1 (see Fig.\;\ref{fig:GeomArg}(a-c)), as we explain.  We desire rhombi-slits and each such slit will have only one free side length, rather than the two free lengths of a generic parallelogram. This restricts the seed: its second quad \textit{must} be a reflection of its first quad across the vector $\mathbf{W}\mathbf{t}$, to obtain rhombi slits. Since the seed quads are mirrored, the $180^{\circ}$ rotation from before is now a reflection across $\mathbf{t}$. Hence, $\mathbf{s}$ is parallel to $\mathbf{W} \mathbf{t}$, i.e., the lattice vectors $\mathbf{s}$ and $\mathbf{t}$ are orthogonal.  Without loss of generality, we can rotate and rescale the pattern to make $\mathbf{s}$ into $\mathbf{e}_1$. It follows that $\mathbf{t} = a_r \mathbf{e}_2$ for some $a_r >0$. We then obtain a unit cell exactly as described in Fig.\;3, consistent with (\ref{eq:sVtVExplicit}). 

\section{Coarse-graining soft modes}
In this section, we derive the effective PDE in Eq.\;(3) by constructing a general, locally mechanistic soft mode. The idea is to allow spatial variations in the mechanisms of the previous section; an effective deformation $\mathbf{y}_\text{eff}(\mathbf{x})$ emerges in the bulk. As stated in the main text, we find that $\mathbf{y}_\text{eff}(\mathbf{x})$ describes a soft mode when it solves the PDE. We enunciate a basic numerical procedure based on our construction to find the panel motions. It is used in the theory portions of Fig.\;4 of the main text. We end with a discussion of the discrepancies between theory and experiment in our approach.

\subsection{Closing the gaps in a soft mode}\label{s:Gaps}

\begin{figure}[t]
\centering
\includegraphics[width =\linewidth]{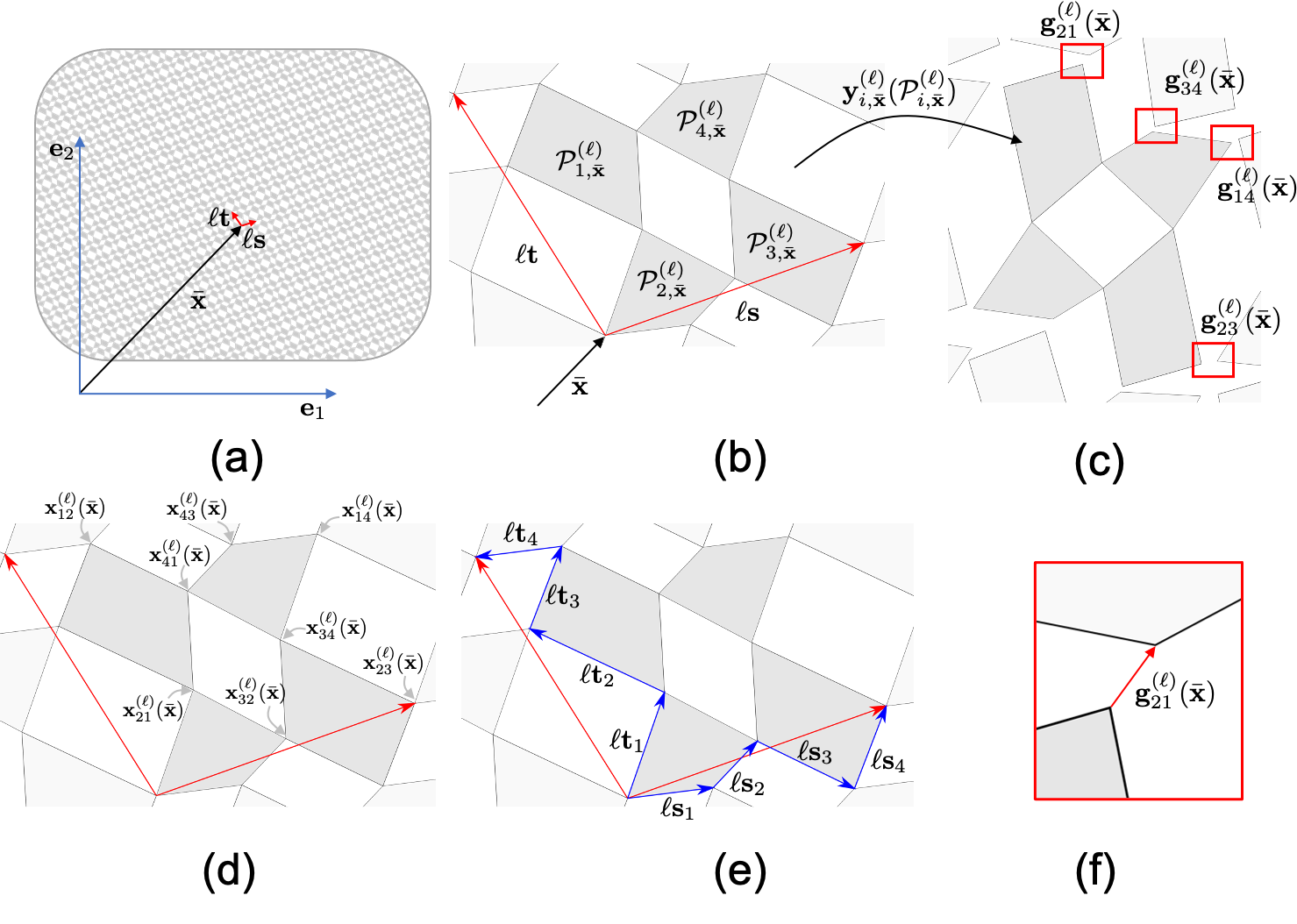}
\caption{Notation for deriving the effective description and corresponding simulation method.}
\label{fig:Ansatz}
\end{figure}

We start by fixing our notation with reference to Fig.\;\ref{fig:Ansatz}. Fix a general periodic quad-kirigami pattern as in Section \ref{s:UC}, whose unit cell has four convex quad panels and four parallelogram slits. The system we study is made up of a large but finite number of kirigami unit cells selected from this pattern, within a two-dimensional reference domain $\Omega\subset\mathbb{R}^2$ (Fig.\;\ref{fig:Ansatz}(a)).
Non-dimensionalizing by the system size, we take our panels and unit cells to have (non-dimensional) widths $\sim \ell \ll 1$. Each cell is labeled by its lower left corner point, which we consistently write with an overbar, as in $\bar{\mathbf{x}}$. The four panels in the cell at $\bar{\mathbf{x}}$ are then labeled counter-clockwise as $\mathcal{P}_{1, \bar{\mathbf{x}}}^{(\ell)}, \ldots, \mathcal{P}_{4,\bar{\mathbf{x}}}^{(\ell)}$ (Fig.\;\ref{fig:Ansatz}(b)). 

We consider panel deformations made up of counter-rotations and translations in the form
\begin{equation}
\begin{aligned}\label{eq:ansatzRigid}
&\mathbf{y}_{i,\bar{\mathbf{x}}}^{(\ell)}(\mathbf{x}) = \mathbf{R}(\gamma(\bar{\mathbf{x}}) + \sigma_i \xi(\bar{\mathbf{x}})) (\mathbf{x} - \bar{\mathbf{x}}) + \mathbf{y}_{\text{eff}}(\bar{\mathbf{x}}) + \ell \mathbf{d}_i(\bar{\mathbf{x}}),\quad \mathbf{x} \in \mathcal{P}_{i,\bar{\mathbf{x}}}^{(\ell)}
\end{aligned}
\end{equation} 
for $i = 1,\ldots,4$, and where $\sigma_1 = \sigma_3 = 1$ and $\sigma_2 = \sigma_4 = -1$. The effective deformation $\mathbf{y}_\text{eff}(\mathbf{x})$ describes a macroscopic motion we think of as occurring in a smooth manner on the whole domain $\Omega$, even though it is only sampled at the cell reference points $\bar{\mathbf{x}}$. Similarly, we refer to macroscopic angle functions $(\gamma(\mathbf{x}), \xi(\mathbf{x}))$ and panel translation functions $\mathbf{d}_1(\mathbf{x}), \ldots, \mathbf{d}_4(\mathbf{x})$.   
In this ansatz, initially connected panels can and generically will have gaps open up between them as in Fig.\;\ref{fig:Ansatz}(c,f). We will choose the panel translations $\mathbf{d}_i(\mathbf{x})$ to close the gaps within each deformed unit cell, thus enforcing a locally mechanistic response. Then, the only gaps that occur are between neighboring unit cells (Fig.\;\ref{fig:Ansatz}(c)). As we shall see, these gaps are consistent with a soft mode (and, in particular, are of order $\sim\ell^2$) provided $\mathbf{y}_\text{eff}(\mathbf{x})$ solves the effective PDE.

To help organize the calculation, we label the corner points of the $\bar{\mathbf{x}}$-cell as $\mathbf{x}_{ij}^{(\ell)}(\bar{\mathbf{x}})$ (Fig.\;\ref{fig:Ansatz}(d)). We then follow the calculation from section \ref{sec:KinAppend} using the replacements 
$\mathbf{R}_i \mapsto \mathbf{R}_i(\bar{\mathbf{x}})$, $\mathbf{c}_i \mapsto \mathbf{y}_{\text{eff}}(\bar{\mathbf{x}}) - \mathbf{R}_i(\bar{\mathbf{x}})\bar{\mathbf{x}} + \ell \mathbf{d}_{i}(\bar{\mathbf{x}})$ and $\mathbf{x}_{ij} \mapsto \mathbf{x}_{ij}^{(\ell)}(\bar{\mathbf{x}})$ to compare (\ref{eq:ansatzRigid}) and (\ref{eq:Rigid0}).
Copying the result of (\ref{eq:c123}), we see that the $\bar{\mathbf{x}}$-cell remains gap-free if and only if 
\begin{equation}
\begin{aligned}\label{eq:ciResults}
\ell ( \mathbf{d}_1(\bar{\mathbf{x}}) - \mathbf{d}_4(\bar{\mathbf{x}})) = \Big( \mathbf{R}(\gamma(\bar{\mathbf{x}}) + \sigma_4 \xi(\bar{\mathbf{x}})) -  \mathbf{R}(\gamma(\bar{\mathbf{x}}) + \sigma_1 \xi(\bar{\mathbf{x}})) \Big) \big( \mathbf{x}_{41}^{(\ell)}(\bar{\mathbf{x}})  - \bar{\mathbf{x}}\big), \\
\ell ( \mathbf{d}_2(\bar{\mathbf{x}}) - \mathbf{d}_1(\bar{\mathbf{x}})) = \Big( \mathbf{R}(\gamma(\bar{\mathbf{x}}) + \sigma_1 \xi(\bar{\mathbf{x}})) -  \mathbf{R}(\gamma(\bar{\mathbf{x}}) + \sigma_2 \xi(\bar{\mathbf{x}})) \Big) \big( \mathbf{x}_{12}^{(\ell)}(\bar{\mathbf{x}})  - \bar{\mathbf{x}}\big), \\
\ell ( \mathbf{d}_3(\bar{\mathbf{x}}) - \mathbf{d}_2(\bar{\mathbf{x}})) = \Big( \mathbf{R}(\gamma(\bar{\mathbf{x}}) + \sigma_2 \xi(\bar{\mathbf{x}})) -  \mathbf{R}(\gamma(\bar{\mathbf{x}}) + \sigma_3 \xi(\bar{\mathbf{x}})) \Big) \big( \mathbf{x}_{32}^{(\ell)}(\bar{\mathbf{x}})  - \bar{\mathbf{x}}\big).
\end{aligned}
\end{equation}
There are six constraints for the eight degrees of freedom in question. Following Fig.\;\ref{fig:Ansatz}(d,e), we have 
\begin{equation}
\begin{aligned}\label{eq:xijParams}
\mathbf{x}^{(\ell)}_{41}(\bar{\mathbf{x}}) = \bar{\mathbf{x}} + \ell(\mathbf{t}_1 + \mathbf{t}_2 + \mathbf{t}_3 + \mathbf{s}_3), \quad \mathbf{x}^{(\ell)}_{12}(\bar{\mathbf{x}}) = \bar{\mathbf{x}} + \ell \mathbf{t}_1, \quad \mathbf{x}^{(\ell)}_{32}(\bar{\mathbf{x}}) = \bar{\mathbf{x}} + \ell ( \mathbf{s}_1 + \mathbf{s}_2).
\end{aligned}
\end{equation}
Combining (\ref{eq:ciResults}) and (\ref{eq:xijParams}) gives the constraints 
\begin{equation}
\begin{aligned}\label{eq:ciPrescribe}
&\mathbf{d}_1(\mathbf{x}) = \Big( \mathbf{R}(\gamma(\mathbf{x}) - \xi(\mathbf{x}) )-  \mathbf{R}(\gamma(\mathbf{x}) +  \xi(\mathbf{x})) \Big) \big( \mathbf{t}_1 + \mathbf{t}_2 + \mathbf{t}_3  + \mathbf{s}_3\big) + \mathbf{d}_4(\mathbf{x}), \\
&\mathbf{d}_2(\mathbf{x}) = \Big( \mathbf{R}(\gamma(\mathbf{x}) - \xi(\mathbf{x}) )-  \mathbf{R}(\gamma(\mathbf{x}) +  \xi(\mathbf{x})) \Big) \big(  \mathbf{t}_2 + \mathbf{t}_3 + \mathbf{s}_3\big) + \mathbf{d}_4(\mathbf{x}) , \\
&\mathbf{d}_3(\mathbf{x}) = \Big( \mathbf{R}(\gamma(\mathbf{x}) - \xi(\mathbf{x}) )-  \mathbf{R}(\gamma(\mathbf{x}) +  \xi(\mathbf{x})) \Big) \big(  \mathbf{t}_2 + \mathbf{t}_3 + \mathbf{s}_1 + \mathbf{s}_2 +  \mathbf{s}_3\big)  + \mathbf{d}_4(\mathbf{x}).
\end{aligned}
\end{equation}
Again, these equations ensure that the panels within each unit cell remain connected upon deformation. Enforcing them prescribes three of the panel translation fields $\mathbf{d}_i(\mathbf{x})$, but not all four.

Next, we attend to the gaps between unit cells, which are given by
\begin{equation}
\begin{aligned}\label{eq:gapsDef}
&\mathbf{g}_{14}^{(\ell)}(\bar{\mathbf{x}}) = \mathbf{y}^{(\ell)}_{1, \bar{\mathbf{x}} + \ell \mathbf{s}} (\mathbf{x}^{(\ell)}_{14}(\bar{\mathbf{x}}))  -  \mathbf{y}^{(\ell)}_{4, \bar{\mathbf{x}}} (\mathbf{x}^{(\ell)}_{14}(\bar{\mathbf{x}})), \quad 
\mathbf{g}_{23}^{(\ell)}(\bar{\mathbf{x}}) = \mathbf{y}^{(\ell)}_{2, \bar{\mathbf{x}} + \ell \mathbf{s}} (\mathbf{x}^{(\ell)}_{23}(\bar{\mathbf{x}}))  - \mathbf{y}^{(\ell)}_{3, \bar{\mathbf{x}}}(\mathbf{x}^{(\ell)}_{23}(\bar{\mathbf{x}})) , \\
&\mathbf{g}_{21}^{(\ell)}(\bar{\mathbf{x}}) = \mathbf{y}^{(\ell)}_{2, \bar{\mathbf{x}} + \ell \mathbf{t}} (\mathbf{x}^{(\ell)}_{21}(\bar{\mathbf{x}}))  -  \mathbf{y}^{(\ell)}_{1, \bar{\mathbf{x}}} (\mathbf{x}^{(\ell)}_{21}(\bar{\mathbf{x}})), \quad \mathbf{g}_{34}^{(\ell)}(\bar{\mathbf{x}}) = \mathbf{y}^{(\ell)}_{3, \bar{\mathbf{x}} + \ell \mathbf{t}} (\mathbf{x}^{(\ell)}_{34}(\bar{\mathbf{x}}))  -  \mathbf{y}^{(\ell)}_{4, \bar{\mathbf{x}}} (\mathbf{x}^{(\ell)}_{34}(\bar{\mathbf{x}}))
\end{aligned}
\end{equation}
as in Fig.\;\ref{fig:Ansatz}(c,d,f). An inspired Taylor expansion will lead to the PDE for $\mathbf{y}_\text{eff}(\mathbf{x})$. Recall the deformed $\bar{\mathbf{x}}$-cell remains connected, by (\ref{eq:ciPrescribe}). Copy it twice and translate the copies along $\ell \mathbf{R}(\gamma(\bar{\mathbf{x}})) \mathbf{A}(\xi(\bar{\mathbf{x}})) \mathbf{s}$ and $\ell \mathbf{R}(\gamma(\bar{\mathbf{x}})) \mathbf{A}(\xi(\bar{\mathbf{x}})) \mathbf{t}$ to produce a perfectly connected ``reference mechanism'' (recall in Fig.\;2, $\mathbf{s}_{\text{def}} = \mathbf{R}(\gamma) \mathbf{A}(\xi)\mathbf{s}$ and $\mathbf{t}_{\text{def}} = \mathbf{R}(\gamma) \mathbf{A}(\xi) \mathbf{t}$). Using this reference mechanism to compare, we see that 
\begin{equation}
\begin{aligned}\label{eq:firstAwesome}
&\mathbf{y}^{(\ell)}_{4, \bar{\mathbf{x}}} (\mathbf{x}^{(\ell)}_{14}(\bar{\mathbf{x}})) = \mathbf{y}^{(\ell)}_{1, \bar{\mathbf{x}}}( \mathbf{x}^{(\ell)}_{14}(\bar{\mathbf{x}})) + \ell  \mathbf{R}(\gamma(\bar{\mathbf{x}})) \mathbf{A}(\xi(\bar{\mathbf{x}})) \mathbf{s}, 
 &\mathbf{y}^{(\ell)}_{3, \bar{\mathbf{x}}} (\mathbf{x}^{(\ell)}_{23}(\bar{\mathbf{x}})) = \mathbf{y}^{(\ell)}_{2, \bar{\mathbf{x}}}(\mathbf{x}^{(\ell)}_{23}(\bar{\mathbf{x}})) + \ell  \mathbf{R}(\gamma(\bar{\mathbf{x}})) \mathbf{A}(\xi(\bar{\mathbf{x}})) \mathbf{s}, \\
 &\mathbf{y}^{(\ell)}_{1, \bar{\mathbf{x}}} (\mathbf{x}^{(\ell)}_{21}(\bar{\mathbf{x}})) = \mathbf{y}^{(\ell)}_{2, \bar{\mathbf{x}}}( \mathbf{x}^{(\ell)}_{21}(\bar{\mathbf{x}})) + \ell  \mathbf{R}(\gamma(\bar{\mathbf{x}})) \mathbf{A}(\xi(\bar{\mathbf{x}})) \mathbf{t}, 
 &\mathbf{y}^{(\ell)}_{4, \bar{\mathbf{x}}} (\mathbf{x}^{(\ell)}_{34}(\bar{\mathbf{x}})) = \mathbf{y}^{(\ell)}_{3, \bar{\mathbf{x}}}(\mathbf{x}^{(\ell)}_{34}(\bar{\mathbf{x}})) + \ell  \mathbf{R}(\gamma(\bar{\mathbf{x}})) \mathbf{A}(\xi(\bar{\mathbf{x}})) \mathbf{t}.
\end{aligned}
\end{equation}
On the other hand by Taylor expansion, 
\begin{equation}
\begin{aligned}\label{eq:secondAwesome}
\mathbf{y}^{(\ell)}_{i,\bar{\mathbf{x}} + \ell \mathbf{s}}(\mathbf{x}) = \mathbf{y}^{(\ell)}_{i,\bar{\mathbf{x}}}(\mathbf{x}) +  \ell \nabla \mathbf{y}_{\text{eff}}(\bar{\mathbf{x}}) \mathbf{s}  + O(\ell^2), \\
\mathbf{y}^{(\ell)}_{i,\bar{\mathbf{x}} + \ell \mathbf{t}}(\mathbf{x}) = \mathbf{y}^{(\ell)}_{i,\bar{\mathbf{x}}}(\mathbf{x}) +  \ell \nabla \mathbf{y}_{\text{eff}}(\bar{\mathbf{x}}) \mathbf{t}  + O(\ell^2)
\end{aligned}
\end{equation}
for any $\mathbf{x} \in \mathcal{P}_{i,\bar{\mathbf{x}}}^{(\ell)}$. Substituting (\ref{eq:firstAwesome}) and (\ref{eq:secondAwesome}) into (\ref{eq:gapsDef}), we conclude the following formulas for the gaps between unit cells:  
\begin{equation}
\begin{aligned}\label{eq:gapsCalc}
&\mathbf{g}_{14}^{(\ell)}(\bar{\mathbf{x}})  = \ell\big(\nabla \mathbf{y}_{\text{eff}}(\bar{\mathbf{x}}) - \mathbf{R}(\gamma(\bar{\mathbf{x}})) \mathbf{A}(\xi(\bar{\mathbf{x}}))  \big) \mathbf{s}  + O(\ell^2), 
&\mathbf{g}_{23}^{(\ell)}(\bar{\mathbf{x}})  = \ell\big(\nabla \mathbf{y}_{\text{eff}}(\bar{\mathbf{x}}) - \mathbf{R}(\gamma(\bar{\mathbf{x}})) \mathbf{A}(\xi(\bar{\mathbf{x}}))  \big) \mathbf{s}  + O(\ell^2), \\
&\mathbf{g}_{21}^{(\ell)}(\bar{\mathbf{x}})  = \ell\big(\nabla \mathbf{y}_{\text{eff}}(\bar{\mathbf{x}}) - \mathbf{R}(\gamma(\bar{\mathbf{x}})) \mathbf{A}(\xi(\bar{\mathbf{x}}))  \big) \mathbf{t}  + O(\ell^2), 
&\mathbf{g}_{34}^{(\ell)}(\bar{\mathbf{x}})  = \ell\big(\nabla \mathbf{y}_{\text{eff}}(\bar{\mathbf{x}}) - \mathbf{R}(\gamma(\bar{\mathbf{x}})) \mathbf{A}(\xi(\bar{\mathbf{x}}))  \big) \mathbf{t}  + O(\ell^2).
\end{aligned}
\end{equation}
These formulas prescribe the gaps between the unit cell at $\bar{\mathbf{x}}$ and its neighbors above and to the right (Fig.\;\ref{fig:Ansatz}(c)). Letting $\bar{\mathbf{x}}$ vary throughout the unit cells recovers all the gaps. Thus, the gaps in our prescription (\ref{eq:ansatzRigid}) are of order $\sim \ell^2$ if and only if the effective PDE holds:
\begin{equation}
\begin{aligned}\label{eq:effectDescribe}
\nabla \mathbf{y}_{\text{eff}}(\mathbf{x}) = \mathbf{R}(\gamma(\mathbf{x})) \mathbf{A}(\xi(\mathbf{x})).
\end{aligned}
\end{equation}

\subsection{A numerical procedure for plotting panel motions}\label{s:Simulations}
The preceding description can be used to visualize the panel motions of a soft mode. Given a solution $(\mathbf{y}_{\text{eff}}(\mathbf{x}), \gamma(\mathbf{x}), \xi(\mathbf{x}))$ of the effective PDE (\ref{eq:effectDescribe}), we substitute it and the formulas for each $\mathbf{d}_i(\mathbf{x})$ from (\ref{eq:ciPrescribe}) into the rigid panel motion ansatz from (\ref{eq:ansatzRigid}). To generate the theoretical results in Fig.\;4, we take $\mathbf{d}_4(\mathbf{x}) = \mathbf{0}$, though in general this choice could be optimized to improve the gaps at order $\sim \ell^2$. (That this optimization does not improve the gaps at a lower order follows from (\ref{eq:gapsCalc}).)  By this choice, each panel motion $\mathbf{y}_{i,\bar{\mathbf{x}}}^{(\ell)}(\mathbf{x})$ in (\ref{eq:effectDescribe}) is fully assigned: we achieve a recipe for deforming all panels in the pattern by rotations and translations approximating the effective deformation $\mathbf{y}_{\text{eff}}(\mathbf{x})$. The gaps between unit cells are at most $O(\ell^2)$, and are negligible compared to the unit cell widths in the limit of a large number of cells. 

\subsection{Elastic energy}

Having evaluated the gaps in our ansatz for the panel motions, we now determine when the construction yields a \textit{soft mode}, which by definition has stored elastic energy much smaller than bulk elasticity scaling. We do so by introducing a simple spring model for the energy, and coarse-graining it in the doubly asymptotic limit mentioned in the main text.  The effective PDE (\ref{eq:effectDescribe}) emerges to set the bulk energy; we conclude it must  be solved for our ansatz to generate a soft mode.

Consider the ansatz in (\ref{eq:ansatzRigid}-\ref{eq:gapsCalc}), this time without necessarily enforcing the PDE for $\mathbf{y}_\text{eff}(\mathbf{x})$. While the construction is purely geometric, one can enrich it with  distortions that \textit{elastically close the gaps}. Two sources of energy result: the panels and hinges (i) stretch to preserve the pattern's topology, and (ii) bend to account for the counter-rotations. Adopting a simple elastic spring model as in \cite{Deng2020}, we define the energy
\begin{equation}
    \begin{aligned}
    \mathcal{E}_{\text{SM}}^{(\ell,\delta)} = \sum_{\bar{\mathbf{x}}} \sum_{i\sim j} \frac{1}{2}E \Big( |\mathbf{g}_{ij}^{(\ell)}(\bar{\mathbf{x}})|^2  + \delta^2 w_{ij}(\xi(\bar{\mathbf{x}})) \Big)  
    \end{aligned}
\end{equation}
under the rigid ansatz (\ref{eq:ansatzRigid}). The first term describes a linear spring with zero rest length linking the corner points of adjacent panels; it accounts for the forces needed to close gaps of length $|\mathbf{g}_{ij}^{(\ell)}(\bar{\mathbf{x}})|$.  The second term describes a (possibly nonlinear) torsional spring energy $w_{ij}(\xi)$, and accounts for the bending moments in the hinges.  Per Fig.\;\ref{fig:Fab}(b), the experimental patterns come with hinges of length $d \approx (1/10)\ell$ and height $h \approx (1/40)\ell$; in the model, we lump these dimensions into a single characteristic length $\delta \ll \ell$ and take the hinge energy to be $\sim \delta^2$. Consistent with our experiments in which the panels and hinges are made of the same material, we let $E >0$ denote a single characteristic elastic modulus. As usual, we use $\bar{\mathbf{x}}$ to denote a choice of unit cell in the  reference domain $\Omega$. The inner sum over $i\sim j$ then collects the spring energies associated to the eight vertices of the $\bar{\mathbf{x}}$-cell shown in Fig.\;\ref{fig:Ansatz}(d). 

Having defined an energy, we coarse-grain it in a limit 
\begin{equation}
    \begin{aligned}\label{eq:asymtotics}
    \ell \rightarrow 0 \quad \text{with $\delta \equiv \delta(\ell)$ chosen to satisfy} \quad \frac{\delta(\ell)}{\ell} \rightarrow 0.
    \end{aligned}
\end{equation}
This requires that the hinges remain asymptotically negligible in size relative to the panels, while the number of panels grows. We additionally assume that the quantities $\mathbf{y}_{\text{eff}}(\mathbf{x}), \gamma(\mathbf{x})$ and $\xi(\mathbf{x})$ from our ansatz are chosen independently of $\ell$ (see, however, \ref{sec:Comparison} for a brief discussion of how to relax this constraint). By the formulas for the gaps in (\ref{eq:gapsCalc}) and the definitions 
\begin{equation}
   w(\xi):= \sum_{i \sim j} w_{ij}(\xi)  \quad\text{and}\quad Q_{\text{bulk}}(\mathbf{G}) :=2\Tr\Big( \mathbf{G}^T\mathbf{G} ( \mathbf{s} \otimes \mathbf{s} + \mathbf{t} \otimes \mathbf{t})\Big),
\end{equation} 
we have that
\begin{equation}
    \begin{aligned}
    \mathcal{E}_{\text{SM}}^{(\ell,\delta(\ell))} &=\sum_{\bar{\mathbf{x}}}E \Big(|\mathbf{g}_{21}^{(\ell)}(\bar{\mathbf{x}})|^2 + |\mathbf{g}_{34}^{(\ell)}(\bar{\mathbf{x}})|^2  + |\mathbf{g}_{14}^{(\ell)}(\bar{\mathbf{x}})|^2 + |\mathbf{g}_{23}^{(\ell)}(\bar{\mathbf{x}})|^2 + \delta^2 w(\xi(\bar{\mathbf{x}})) \Big)  \\
    &= \sum_{\bar{\mathbf{x}}}E \Big\{2\ell^2  \sum_{\mathbf{v} \in \{\mathbf{s},\mathbf{t}\}}|\big(\nabla \mathbf{y}_{\text{eff}}(\bar{\mathbf{x}}) - \mathbf{R}(\gamma(\bar{\mathbf{x}})) \mathbf{A}(\xi(\bar{\mathbf{x}}))  \big) \mathbf{v} + O(\ell)|^2  + \delta^2 w(\xi(\bar{\mathbf{x}})) \Big\} \\
    &=\sum_{\bar{\mathbf{x}}} \ell^2 E \Big\{  Q_{\text{bulk}}\Big(\nabla \mathbf{y}_{\text{eff}}(\bar{\mathbf{x}}) - \mathbf{R}(\gamma(\bar{\mathbf{x}})) \mathbf{A}(\xi(\bar{\mathbf{x}}))  \Big) + \frac{\delta^2}{\ell^2}w(\xi(\bar{\mathbf{x}})) + O(\ell) \Big\}.
    \end{aligned}
\end{equation}
Recognize, in view of Fig.\;\ref{fig:Ansatz}(a-b), that each unit cell covers an area  $\ell^2 |\mathbf{s} \cdot \mathbf{W} \mathbf{t}|$. Applying the Riemann integration formula   $\sum_{\bar{\mathbf{x}}}\ell^2 f(\bar{\mathbf{x}}) \to \frac{1}{|\mathbf{s} \cdot \mathbf{W} \mathbf{t}|}\int_{\Omega} f(\mathbf{x})\, dA$ yields for the coarse-grained energy
\begin{equation}
    \begin{aligned}\label{eq:takeLimit}
    \lim_{\ell \rightarrow 0}\, \mathcal{E}_{\text{SM}}^{(\ell,\delta(\ell))} &= \lim_{\ell \rightarrow 0}\, \frac{E}{|\mathbf{s} \cdot \mathbf{W}\mathbf{t}|} \int_{\Omega}  \Big\{ Q_{\text{bulk}}\Big(\nabla \mathbf{y}_{\text{eff}}(\mathbf{x}) - \mathbf{R}(\gamma(\mathbf{x})) \mathbf{A}(\xi(\mathbf{x}))  \Big) + \frac{\delta(\ell)^2}{\ell^2} w(\xi(\mathbf{x})) +O(\ell)\Big\}\, dA  \\
    &= \frac{E}{|\mathbf{s} \cdot \mathbf{W}\mathbf{t}|}\int_{\Omega} Q_{\text{bulk}}\Big(\nabla \mathbf{y}_{\text{eff}}(\mathbf{x}) - \mathbf{R}(\gamma(\mathbf{x})) \mathbf{A}(\xi(\mathbf{x}))  \Big)\,dA.
    \end{aligned}
\end{equation}

The bulk elastic energy just derived vanishes if and only if the effective PDE (\ref{eq:effectDescribe}) holds.  This explains why the effective PDE captures the soft modes of our kirigami patterns. When specialized to conformal kirigami, (\ref{eq:takeLimit}) recovers the bulk part of the elastic energy from \cite{czajkowski2022conformal}, up to a choice of elastic moduli. 

\subsection{Comparing experimental soft modes to the coarse-grained theory}\label{sec:Comparison}
The effective PDE  (\ref{eq:effectDescribe}) is fundamentally an asymptotic constraint. It describes the leading order behaviors of soft kirigami modes in the limit of a large number of panels and with negligible hinges (per (\ref{eq:asymtotics})). At small but finite $\ell$ and $\delta/\ell$, the actual deformation of the kirigami can be expected to exhibit slight deviations from our ansatz. We address these deviations now. 

Consider, for instance, a refinement of the panel motions in the ansatz (\ref{eq:ansatzRigid}) obtained by replacing the $\ell$-independent fields $(\mathbf{y}_{\text{eff}}(\mathbf{x}) , \gamma(\mathbf{x}), \xi(\mathbf{x}))$ with ones that are allowed to depend on the number of panels and perhaps even the hinge sizes:
\begin{equation}
    \begin{aligned}\label{eq:perturb}
    \mathbf{y}_{\text{eff}}(\mathbf{x}) &\mapsto \mathbf{y}_{\text{eff}}^{(\ell,\delta)}(\mathbf{x}) = \mathbf{y}_{0}(\mathbf{x}) + \ell \mathbf{y}_1(\mathbf{x}) + \frac{\delta^2}{\ell^2} \mathbf{y}_2(\mathbf{x}),  \\
    \mathbf{\gamma}(\mathbf{x}) &\mapsto \gamma^{(\ell,\delta)}(\mathbf{x}) = \gamma_0(\mathbf{x}) + \ell \gamma_1(\mathbf{x}) + \frac{\delta^2}{\ell^2} \gamma_2(\mathbf{x}), \\ 
    \xi(\mathbf{x}) &\mapsto \xi^{(\delta,\ell)}(\mathbf{x}) = \xi_0(\mathbf{x}) + \ell \xi_1(\mathbf{x}) + \frac{\delta^2}{\ell^2} \xi_2(\mathbf{x}).
    \end{aligned}
\end{equation}
After updating the previous asymptotic analysis to account for these replacements, one finds a coarse-grained energy nearly identical to that of (\ref{eq:takeLimit}), the only difference being that the term inside $Q_{\text{bulk}}(\cdot)$ is replaced by $\nabla \mathbf{y}_{0}(\mathbf{x}) - \mathbf{R}(\gamma_0(\mathbf{x})) \mathbf{A}(\xi_0(\mathbf{x})).$ Setting $\nabla \mathbf{y}_{0}(\mathbf{x}) = \mathbf{R}(\gamma_0(\mathbf{x})) \mathbf{A}(\xi_0(\mathbf{x}))$ to produce a soft mode leaves room for perturbations in the effective description which now scale with the parameters:
\begin{equation}
\begin{aligned}\label{eq:PDEApprox}
\Big | \nabla \mathbf{y}_{\text{eff}}^{(\ell,\delta)}(\mathbf{x}) - \mathbf{R}(\gamma^{(\ell,\delta)}(\mathbf{x})) \mathbf{A}(\xi^{(\ell,\delta)}(\mathbf{x}))\Big | \sim \max \Big\{ \ell, \frac{\delta^2}{\ell^2} \Big\}.
\end{aligned}
\end{equation}
The choice of the remaining perturbations in (\ref{eq:perturb}) is a delicate matter involving minimizing the residual elastic energy of a general  pattern. The  derivation of this energy is the subject of ongoing research; for more in the case of conformal kirigami, see \cite{czajkowski2022conformal}. 

Even without settling the exact form of the residual energy, we can still address the order of magnitude of the discrepancies between our experiments and theory. The patterns in Fig.\;4 are $16\times16$ unit cells, giving $\ell \sim 1/16\mbox{-}1/32$. Per Fig.\;\ref{fig:Fab},  $\delta^2/\ell^2 \sim \hat{d}\cdot \hat{h} = 1/400$. So, $\ell \gg \delta^2/\ell^2$ in our samples.  By (\ref{eq:PDEApprox}), the deformation gradients in the experiments can reasonably deviate from the effective PDE constraint (\ref{eq:effectDescribe}) on the order of $3\mbox{-}6\%$ of the magnitude of the field variables $\mathbf{y}_{\text{eff}}(\mathbf{x}), \gamma(\mathbf{x})$ and $\xi(\mathbf{x})$ used in our simulations. This expectation is  consistent with the results   in the main text.

\section{The compatibility condition for effective deformations}

The previous sections concerned the derivation of the effective PDE
\begin{equation}\label{eq:effPDE}
    \nabla \mathbf{y}_{\text{eff}}(\mathbf{x}) = \mathbf{R}(\gamma(\mathbf{x})) \mathbf{A}(\xi(\mathbf{x}))
\end{equation}
relating $\mathbf{y}_\text{eff}(\mathbf{x})$ to the angle fields $(\gamma(\mathbf{x}),\xi(\mathbf{x}))$. The rest of this supplement is about the analysis of this PDE and its solutions. We start in this section by deriving the compatibility condition 
\begin{equation}\label{eq:compat1}
\nabla \gamma(\mathbf{x}) = \boldsymbol{\Gamma}(\xi(\mathbf{x})) \nabla \xi(\mathbf{x})  
\end{equation}
as a consequence of (\ref{eq:effPDE}). Actually, on simply connected domains $\Omega$, this condition is not only necessary but also sufficient for the fields $(\gamma(\mathbf{x}),\xi(\mathbf{x}))$ to admit some effective deformation $\mathbf{y}_\text{eff}(\mathbf{x})$ solving (\ref{eq:effPDE}). Indeed, we shall derive it by enforcing the curl-free nature of $\nabla\mathbf{y}_\text{eff}(\mathbf{x})$. Sufficiency follows since a curl-free tensor field on a simply connected domain is always the gradient of a vector field.  

Let $\mathbf{y}_{\text{eff}}(\mathbf{x})$, $\gamma(\mathbf{x})$ and $\xi(\mathbf{x})$ satisfy the effective PDE (\ref{eq:effPDE}) on $\Omega\subset\mathbb{R}^2$. Since partial derivatives  commute, i.e., $\partial_1\partial_2 \mathbf{y}_{\text{eff}}(\mathbf{x}) = \partial_2 \partial_1 \mathbf{y}_{\text{eff}}(\mathbf{x})$,
\begin{equation}
\begin{aligned}\label{eq:curlFree}
\partial_1\Big( \mathbf{R}(\gamma(\mathbf{x})) \mathbf{A}(\xi(\mathbf{x})) \mathbf{e}_2\Big) - \partial_2 \Big( \mathbf{R}(\gamma(\mathbf{x})) \mathbf{A}(\xi(\mathbf{x})) \mathbf{e}_1 \Big) = \mathbf{0}.
\end{aligned}
\end{equation}
Using the chain and product rules, there follows
\begin{equation}
\begin{aligned}\label{eq:chainRule}
\mathbf{R}(\gamma(\mathbf{x})) \Big[ \partial_1 \gamma(\mathbf{x}) \mathbf{W}  \mathbf{A}(\xi(\mathbf{x})) \mathbf{e}_2 +  \partial_1 \xi(\mathbf{x})   \mathbf{A}'(\xi(\mathbf{x})) \mathbf{e}_2 -  \partial_2 \gamma(\mathbf{x}) \mathbf{W}  \mathbf{A}(\xi(\mathbf{x})) \mathbf{e}_1 -   \partial_2 \xi(\mathbf{x})   \mathbf{A}'(\xi(\mathbf{x})) \mathbf{e}_1\Big]  = \mathbf{0}
\end{aligned}
\end{equation}
where $\mathbf{W} = \mathbf{R}(\pi/2)$. Note  $\partial_1 \gamma(\mathbf{x}) \mathbf{e}_2 - \partial_2 \gamma(\mathbf{x}) \mathbf{e}_1 = \mathbf{W} \nabla \gamma(\mathbf{x})$ and  $\partial_1 \xi(\mathbf{x}) \mathbf{e}_2 - \partial_2 \xi(\mathbf{x}) \mathbf{e}_1 = \mathbf{W} \nabla \xi(\mathbf{x})$.  Hence, (\ref{eq:chainRule}) re-writes as 
\begin{equation}
\begin{aligned}\label{eq:firstRewrite}
\mathbf{W} \mathbf{A}(\xi(\mathbf{x})) \mathbf{W} \nabla \gamma(\mathbf{x}) + \mathbf{A}'(\xi(\mathbf{x})) \mathbf{W} \nabla \xi(\mathbf{x}) = \mathbf{0},
\end{aligned}
\end{equation}
after pre-multiplying by $\mathbf{R}^T(\gamma(\mathbf{x}))$. Writing $\mathbf{W} \mathbf{A}(\xi) \mathbf{W}$ in the Cartesian basis shows that 
\begin{equation}
    \begin{aligned}\label{eq:coolIdent}
    -\mathbf{W} \mathbf{A}(\xi) \mathbf{W} = -\left(\begin{array}{cc} 0 & -1 \\ 
    1 & 0  \end{array}\right) \left(\begin{array}{cc} A_{11}(\xi) & A_{12}(\xi) \\ 
    A_{21}(\xi) & A_{22}(\xi)  \end{array}\right)\left(\begin{array}{cc} 0 & -1 \\ 
    1 & 0  \end{array}\right) = \left(\begin{array}{cc} A_{22}(\xi) & -A_{21}(\xi) \\ 
    -A_{12}(\xi) & A_{11}(\xi)  \end{array}\right) := \text{cof} \mathbf{A}(\xi)
    \end{aligned}
\end{equation}
for the \textit{cofactor} matrix of $\mathbf{A}(\xi)$.
Of course, $\det \nabla \mathbf{y}_{\text{eff}}(\mathbf{x})  = \det \mathbf{A}(\xi(\mathbf{x}))>0$ in all physically relevant cases. Then,  $\text{cof }\mathbf{A}(\xi) = \big(\det \mathbf{A}(\xi) \big) \mathbf{A}^{-T}(\xi)$, where $\mathbf{A}^{-T}(\xi)$ is the well-defined inverse transpose of $\mathbf{A}(\xi)$. By making use of all these identities, the equation in (\ref{eq:firstRewrite}) becomes
\begin{equation}
\begin{aligned}\label{eq:finalCompat}
\nabla \gamma(\mathbf{x}) =\underbrace{\Big[  \frac{\mathbf{A}^T(\xi(\mathbf{x})) \mathbf{A}'(\xi(\mathbf{x})) \mathbf{W}}{\det \mathbf{A}(\xi(\mathbf{x}))} \Big]}_{:= \boldsymbol{\Gamma}(\xi(\mathbf{x}))} \nabla \xi(\mathbf{x}),
\end{aligned}
\end{equation}
which is the desired compatibility condition (\ref{eq:compat1}).

\section{PDE type and a universal link to the Poisson's ratio}\label{s:PDEClass}

Here, we classify the general quad-based kirigami patterns from the main text using the concept of PDE type. In particular, we find the type of the compatibility relations relating the angle fields:
\begin{equation}\label{eq:forPDEType}
\nabla \gamma(\mathbf{x}) = \boldsymbol{\Gamma}(\xi(\mathbf{x})) \nabla \xi(\mathbf{x}).  
\end{equation}
Since we are in two dimensions, this system comprises two first order scalar PDEs in the unknowns $(\gamma(\mathbf{x}),\xi(\mathbf{x}))$. It is therefore equivalent to a single second order nonlinear scalar PDE in, say, $\xi(\mathbf{x})$. Following the standard procedure, we determine the type of this equation as \textit{elliptic}, \textit{hyperbolic}, or \textit{parabolic} depending on the coefficients of its linearization about a solution \cite[Ch.\;III; Sec.\;1.3]{courant2008methods}. 

Note in the examples from Fig.\;4, the PDE type remains constant (it is either elliptic or hyperbolic). In general, however, the type can vary across  $\Omega$. We first illustrate this procedure on a mechanism deformation, and then discuss general soft modes. We end by finding the link between the PDE type of (\ref{eq:forPDEType}) and the effective Poisson's ratio of our kirigami, in the general case stated in the main text.

\subsection{Definition of  PDE type}
As recalled above, the PDE type of (\ref{eq:forPDEType}) is defined through  linearization. To help explain this procedure, we first demonstrate it on a pure mechanism. Then, we obtain the type for a general soft mode.

Consider a mechanism deformation given by the constant angle functions $(\gamma_0(\mathbf{x}),\xi_0(\mathbf{x}))\equiv(\gamma_0,\xi_0)$. Substituting the perturbations $\gamma(\mathbf{x}) =  \gamma_0 + \delta \gamma(\mathbf{x})$ and $\xi(\mathbf{x}) = \xi_0 + \delta \xi(\mathbf{x})$ into (\ref{eq:forPDEType}) and collecting terms at leading order yields the linear system
\begin{equation}
    \begin{aligned}
    \nabla \delta \gamma(\mathbf{x}) = \boldsymbol{\Gamma}(\xi_0) \nabla \delta \xi(\mathbf{x}).
    \end{aligned}
\end{equation}
Taking the curl of the lefthand side to eliminate $\delta\gamma(\mathbf{x})$ gives a second order scalar PDE for $\delta\xi(\mathbf{x})$:
\begin{equation}
    \begin{aligned}\label{eq:linearPDESupp}
    \nabla \cdot \Big( \mathbf{W} \boldsymbol{\Gamma}(\xi_0) \nabla \delta \xi(\mathbf{x})\Big)= 0.
    \end{aligned}
\end{equation}
Introducing  $\text{sym}\, \mathbf{A} = \frac{1}{2}(\mathbf{A} + \mathbf{A}^T)$, this becomes 
\begin{equation}
    \begin{aligned}\label{eq:linPDE}
    \big(\sym (\mathbf{W}\boldsymbol{\Gamma}(\xi_0))\big)_{ij} \partial_i \partial_j \delta \xi(\mathbf{x}) = 0
    \end{aligned}
\end{equation}
in index notation with repeated indices summed. Eq.\;(\ref{eq:linPDE}) gives the linearized PDE in standard form. 
Its type is determined via the discriminant of its coefficients \cite{courant2008methods}. Equivalently, it is determined by 
\begin{equation}
    \begin{aligned}\label{eq:sigmaPDE}
    \sigma_{\text{PDE}}(\xi_0) := \det\Big( \text{sym} \big( \mathbf{W} \boldsymbol{\Gamma}(\xi_0)\big)\Big).
    \end{aligned}
\end{equation}
If $\sigma_{\text{PDE}}(\xi_0) >0$, the PDE is said to be of \textit{elliptic} type; if $\sigma_{\text{PDE}}(\xi_0) <0$, it is called \textit{hyperbolic};  if $\sigma_{\text{PDE}}(\xi_0) =0$, it is parabolic. While this classification is made in reference to the linearized PDE (\ref{eq:linPDE}), it is understood to apply to the nonlinear PDE (\ref{eq:forPDEType}) as well.

Returning to generalities, we now linearize the  system (\ref{eq:forPDEType}) about an arbitrary solution $(\gamma_0(\mathbf{x}), \xi_0(\mathbf{x}))$ and determine its type. Taking $\gamma(\mathbf{x}) = \gamma_0(\mathbf{x}) + \delta \gamma(\mathbf{x})$ and $\xi(\mathbf{x}) = \xi_0(\mathbf{x}) + \delta \xi(\mathbf{x})$ gives the linear system
\begin{equation}
    \begin{aligned}\label{eq:linPDE2}
    \nabla \delta \gamma(\mathbf{x}) = \boldsymbol{\Gamma}(\xi_0(\mathbf{x})) \nabla \delta \xi(\mathbf{x}) + \delta \xi(\mathbf{x}) \boldsymbol{\Gamma}'(\xi_0(\mathbf{x})) \nabla \xi_0 (\mathbf{x}) 
    \end{aligned}
\end{equation}
for the angle perturbations $(\delta\gamma(\mathbf{x}),\delta\xi(\mathbf{x}))$. Its highest order terms involve their gradients, and are  the  spatially-varying versions of the analogous terms in (\ref{eq:linPDE}). 
Again, we can take the curl of (\ref{eq:linPDE2}) to eliminate $\delta \gamma(\mathbf{x})$ in favor of $\delta\xi(\mathbf{x})$. The end result is a second order linear scalar PDE 
\begin{equation}
 \begin{aligned}\label{eq:large2ndOrder}
    &c_{11}(\mathbf{x}) \partial_1 \partial_1 \delta \xi(\mathbf{x}) + c_{22}(\mathbf{x}) \partial_2 \partial_2 \delta \xi(\mathbf{x}) + c_{12}(\mathbf{x}) \partial_1 \partial_2 \delta \xi(\mathbf{x}) + b_1(\mathbf{x}) \partial_1 \delta \xi(\mathbf{x}) + b_2(\mathbf{x}) \partial_2 \delta \xi(\mathbf{x}) + a(\mathbf{x}) \delta \xi(\mathbf{x}) = 0
    \end{aligned}
\end{equation}
with coefficients $c_{ij}(\mathbf{x})$, $b_i(\mathbf{x})$ and $a(\mathbf{x})$ depending on  $\xi_0(\mathbf{x})$ and $\nabla \xi_0(\mathbf{x})$. It is classified through its highest order derivatives, which involve the coefficients $c_{ij}(\mathbf{x})$. As a moment's consideration will show, these terms are identical to those of (\ref{eq:linPDE}), except  that the previously constant mechanism angle $\xi_0$ is now replaced by a spatially-varying one $\xi_0(\mathbf{x})$. Thus, the PDE type is determined by  $\sigma_{\text{PDE}}(\xi_0(\mathbf{x}))$ from (\ref{eq:sigmaPDE}). The dependence on $\mathbf{x}$ highlights the possibility that this type may vary throughout $\Omega$.

In summary, the PDE system (\ref{eq:forPDEType}) governing soft modes of our kirigami patterns is classified according to its linearization about a solution $(\gamma_0(\mathbf{x}), \xi_0(\mathbf{x}))$, using the determinant 
\begin{equation}
    \begin{aligned}\label{eq:sigmaPDE-general}
    \sigma_{\text{PDE}}(\xi_0(\mathbf{x})) := \det\Big( \text{sym}\big(\mathbf{W} \boldsymbol{\Gamma}(\xi_0(\mathbf{x}))\big)\Big).
    \end{aligned}
\end{equation}
It is \textit{elliptic} where $\sigma_{\text{PDE}}(\xi_0(\mathbf{x})) >0$, \textit{hyperbolic} where $\sigma_{\text{PDE}}(\xi_0(\mathbf{x})) <0$, and \textit{parabolic} where $\sigma_{\text{PDE}}(\xi_0(\mathbf{x})) =0$. It can be of mixed type in general, but can also be of one type throughout $\Omega$ as demonstrated by the rhombi-slit examples in the main text (see Fig.\;4). 


\subsection{Definition of the effective Poisson's ratio}
Next, we define the effective Poisson's ratio of our general kirigami patterns, including the rhombi-slit ones from the main text as a special case. After identifying the Poison's ratio, we go on to link it to the PDE type in the next section. 

We require a formula for the linear strain of a perturbation. 
As in the main text, we expand about a mechanism using $\mathbf{y}_{\text{eff}}(\mathbf{x}) = \mathbf{A}(\xi_0)\mathbf{x} + \mathbf{u}(\mathbf{A}(\xi_0) \mathbf{x})$, where $\mathbf{u}(\mathbf{y})$ is a small displacement field. It is coupled to the angles $\gamma(\mathbf{x})  = \delta \gamma(\mathbf{x})$ and $\xi(\mathbf{x}) = \xi_0 + \delta \xi(\mathbf{x})$ through the PDE (\ref{eq:effPDE}). (Again, we set $\gamma_0=0$ to remove the free global rotation.) Notice that 
\begin{equation}
    \begin{aligned}
    &\nabla \mathbf{y}_{\text{eff}}(\mathbf{x}) = \mathbf{A}(\xi_0) + \nabla \mathbf{u}(\mathbf{A}(\xi_0) \mathbf{x}) \mathbf{A}(\xi_0),\\
    & \mathbf{R}(\gamma(\mathbf{x})) \mathbf{A}(\xi(\mathbf{x})) = \mathbf{A}(\xi_0) + \delta \gamma(\mathbf{x}) \mathbf{W} \mathbf{A}(\xi_0) + \delta \xi(\mathbf{x}) \mathbf{A}'(\xi_0) 
    \end{aligned}
\end{equation}
to first order in the perturbation.
Since the lefthand sides above are required to be equal by the effective PDE (\ref{eq:effPDE}),   
\begin{equation}
 \begin{aligned}\label{eq:uDisp}
 \nabla \mathbf{u}(\mathbf{A}(\xi_0)\mathbf{x}) = \delta \gamma(\mathbf{x}) \mathbf{W} + \delta \xi(\mathbf{x}) \mathbf{A}'(\xi_0) \mathbf{A}^{-1}(\xi_0)
 \end{aligned}
\end{equation}
to leading order. The linear strain  $\boldsymbol{\varepsilon}(\mathbf{y}) = \sym \nabla \mathbf{u}(\mathbf{y})$ is then given to leading order by
\begin{equation}
    \begin{aligned}\label{eq:firstAux} 
    \boldsymbol{\varepsilon}(\mathbf{A}(\xi_0)\mathbf{x}) = \delta \xi(\mathbf{x}) \sym \big(\mathbf{A}'(\xi_0) \mathbf{A}^{-1}(\xi_0)\big).
    \end{aligned}
\end{equation}
Since this strain is symmetric, it can be represented in the principle directions of strain space as 
\begin{equation}
    \begin{aligned}\label{eq:spectralAux}
    \boldsymbol{\varepsilon}(\mathbf{A}(\xi_0)\mathbf{x})  = \delta \xi(\mathbf{x}) \Big( \varepsilon_1(\xi_0) \mathbf{v}_1(\xi_0) \otimes \mathbf{v}_1(\xi_0) + \varepsilon_2(\xi_0) \mathbf{v}_2(\xi_0) \otimes \mathbf{v}_2(\xi_0)\Big).
    \end{aligned}
\end{equation}
To fix the notation, we let $\{\mathbf{v}_{1}(\xi_0), \mathbf{v}_{2}(\xi_0)\}$ be a righthanded orthonormal basis with $\mathbf{v}_1(\xi_0) \cdot \mathbf{e}_1 > 0$ and $\mathbf{v}_1(\xi_0) \cdot \mathbf{e}_2 \geq 0$.  
Given all this, we can now define the effective Poisson's ratio of our kirigami patterns as
\begin{equation}
    \begin{aligned}
    \nu_{21}(\xi_0) := - \frac{\varepsilon_2(\xi_0)}{\varepsilon_1(\xi_0)}.
    \end{aligned}
\end{equation}
This makes concrete the brief description of the general Poisson's ratio in the main text, and recovers the definition in the rhombi-slit case as well. Note its value depends on the slit actuation, $\xi_0$.  

Importantly, away from singular points where $\varepsilon_1(\xi_0) = 0$, the Poisson's ratio satisfies 
\begin{equation}
    \begin{aligned}
    \sign( \nu_{21}(\xi_0) ) = -\sign( \varepsilon_1(\xi_0) \varepsilon_2(\xi_0)) = -\sign\Big[\det \Big( \sym\big(\mathbf{A}'(\xi_0) \mathbf{A}^{-1}(\xi_0)\big) \Big) \Big].
    \end{aligned}
\end{equation}
This follows from combining the righthand sides of  (\ref{eq:firstAux}) and (\ref{eq:spectralAux}).  Thus, the kirigami pattern's auxeticity at actuation $\xi_0$ is determined by the sign of  
\begin{equation}
    \begin{aligned}\label{eq:sigmaAux}
    \sigma_{\text{Aux}}(\xi_0) := \det \Big( \sym\big(\mathbf{A}'(\xi_0) \mathbf{A}^{-1}(\xi_0)\big)\Big).
    \end{aligned}
\end{equation}
When $\sigma_{\text{Aux}}(\xi_0) > 0$, the pattern has a negative Poisson's ratio and is auxetic; when $\sigma_{\text{Aux}}(\xi_0) < 0$, it has a positive Poisson's ratio and a standard, non-auxetic response. 

\subsection{Link between  PDE type and the Poisson's ratio}
We are now in a position to substantiate the claim from the main text that the PDE type of (\ref{eq:forPDEType}) is linked in general to the effective Poisson's ratio of the kirigami.
Formally, we establish the following universal relationship between (\ref{eq:sigmaPDE}) and (\ref{eq:sigmaAux}):
\begin{equation}
    \begin{aligned}\label{eq:universal}
    \sign\big(\sigma_{\text{Aux}}(\xi)\big) =  \sign \big( \sigma_{\text{PDE}}(\xi)\big) \quad \text{when $\det \mathbf{A}(\xi) >0$}.
    \end{aligned}
\end{equation}
Note $\det \nabla \mathbf{y}_{\text{eff}}(\mathbf{x}) = \det \mathbf{A}(\xi(\mathbf{x})) > 0$  for all physically relevant deformations. The identity (\ref{eq:universal}) shows that our kirigami patterns deform auxetically if and only if their effective PDEs are of elliptic type. Conversely, a hyperbolic PDE type corresponds to a standard, non-auxetic Poisson's ratio.  

The desired identity actually follows from a stronger result regarding the eigenvalues of the symmetric tensors in the definitions of $\sigma_\text{Aux}(\xi)$ and $\sigma_\text{PDE}(\xi)$. The fact is that the eigenvalues of $\sym \big( \mathbf{W} \boldsymbol{\Gamma}(\xi) \big)$
 and $\sym \big(\mathbf{A}'(\xi)  \mathbf{A}^{-1}(\xi)\big)$ are of \textit{opposite signs}, provided $\det \mathbf{A}(\xi(\mathbf{x})) > 0$. This is clear in the rhombi-slit case, where the matrices are diagonal (as follows from the formulas in Eq.\;(5) and (6)). But in the general case, it is not obvious. To verify it, we show that the quadratic forms 
\begin{equation}
    \begin{aligned}
    &Q_{\text{PDE}}(\mathbf{v}) = \mathbf{v} \cdot  \sym \big( \mathbf{W} \boldsymbol{\Gamma}(\xi) \big) \mathbf{v}, \\
    &Q_{\text{Aux}}(\mathbf{v}) = \mathbf{v} \cdot  \sym \big(\mathbf{A}'(\xi)  \mathbf{A}^{-1}(\xi)\big) \mathbf{v}
    \end{aligned}
\end{equation}
are negatives of each other, in suitable coordinates. Since the eigenvalues of the matrices in question are nothing other than the maximum and minimum of $Q_\text{PDE}(\mathbf{v})$ and $Q_\text{Aux}(\mathbf{v})$ amongst unit vectors $|\mathbf{v}|=1$ \cite{lax07}, this establishes the result. 

Observe that 
\begin{equation}
    \begin{aligned}
    Q_{\text{PDE}}(\mathbf{v}) = \mathbf{v} \cdot \mathbf{W} \boldsymbol{\Gamma}(\xi) \mathbf{v}, \quad Q_{\text{Aux}}(\mathbf{v}) = \mathbf{v} \cdot \mathbf{A}'(\xi) \mathbf{A}^{-1}(\xi) \mathbf{v}
    \end{aligned}
\end{equation}
since $\mathbf{v} \cdot \mathbf{B} \mathbf{v} = \mathbf{0}$ whenever $\mathbf{B}$ is a skew tensor. Using the first of these, we get that 
\begin{equation}
    \begin{aligned}
    Q_{\text{PDE}}(\mathbf{W}\tilde{\mathbf{v}}) = -\frac{1}{\det \mathbf{A}(\xi)}\tilde{\mathbf{v}}  \cdot \Big( \mathbf{A}^T(\xi) \mathbf{A}'(\xi) \Big) \tilde{\mathbf{v}}
    \end{aligned}
\end{equation}
by the definition of $\boldsymbol{\Gamma}(\xi)$ in (\ref{eq:finalCompat}). Similarly,  \begin{equation}
    \begin{aligned}\label{eq:QAuxFinal}
    Q_{\text{Aux}}\big( \tfrac{\mathbf{A}(\xi) \tilde{\mathbf{v}}}{|\mathbf{A}(\xi) \tilde{\mathbf{v}}|}\big) = \frac{1}{|\mathbf{A}(\xi)\tilde{\mathbf{v}}|^2} \tilde{\mathbf{v}} \cdot \mathbf{A}^T(\xi) \mathbf{A}'(\xi) \tilde{\mathbf{v}} 
    \end{aligned}
\end{equation}
for any $\tilde{\mathbf{v}} \neq \mathbf{0}$. The resulting forms involve the \textit{same} tensor $\mathbf{A}^T(\xi) \mathbf{A}'(\xi)$. So whenever $\det \mathbf{A}(\xi)> 0$, 
\begin{equation}
    \begin{aligned}
    &\sign\Big( \min_{|\mathbf{v}|=1}\, Q_{\text{PDE}}(\mathbf{v})\Big) = -\sign \Big(\min_{|\tilde{\mathbf{v}}|=1}\, \tilde{\mathbf{v}}\cdot \mathbf{A}^T (\xi) \mathbf{A}'(\xi) \tilde{\mathbf{v}} \Big), \\
    &\sign\Big( \min_{|\mathbf{v}|=1}\, Q_{\text{Aux}}(\mathbf{v})\Big) = \sign \Big(\min_{|\tilde{\mathbf{v}}|=1}\, \tilde{\mathbf{v}}\cdot \mathbf{A}^T (\xi) \mathbf{A}'(\xi) \tilde{\mathbf{v}} \Big).
    \end{aligned}
\end{equation}
The same identities hold with maximization in place of minimization. Hence, (\ref{eq:universal}) is proved.

\section{Exact solutions of the effective PDE}

In this section, we provide a detailed description of the  nonlinear PDE solutions behind the theory portions of Fig.\;4 in the main text. The reader may also wish to consult Section \ref{s:Simulations} which presents the numerical method we use to plot the panel motions.

\subsection{Simple wave solutions}
We start by constructing the nonlinear wave response of our hyperbolic rhombi-slit kirigami, under an assumption that its Poisson's ratio $\nu_{21}(\xi)$ is positive on an interval of slit actuation angles $\xi$. These solutions are used to plot the theory half of Fig.\;4(b).

Consider solving the compatibility equations
\begin{equation}\label{eq:compaptHyper}
\nabla \gamma(\mathbf{x}) = \boldsymbol{\Gamma}(\xi(\mathbf{x})) \nabla \xi(\mathbf{x})
\end{equation}
with $\nu_{21}(\xi(\mathbf{x})) > 0$. Since we are dealing with rhombi-slits, $\boldsymbol{\Gamma}(\xi) = \Gamma_{12}(\xi) \mathbf{e}_1 \otimes \mathbf{e}_2 + \Gamma_{21}(\xi) \mathbf{e}_2 \otimes \mathbf{e}_1$ for  $\Gamma_{12}(\xi)  = -\mu_1'(\xi)/\mu_{2}(\xi)$ and $\Gamma_{21}(\xi) = \mu_2'(\xi)/ \mu_1(\xi)$. Its eigenvectors and corresponding eigenvalues are
\begin{equation}
    \label{eq:eigen}
    \mathbf{v}^{\pm}(\xi) = \mathbf{e}_1  \pm \tfrac{\sqrt{\Gamma_{12}(\xi)\Gamma_{21}(\xi)}}{\Gamma_{12}(\xi)}\mathbf{e}_2\quad\text{and}\quad  
   \Gamma^{\pm}(\xi) = \pm \sqrt{\Gamma_{12}(\xi) \Gamma_{21}(\xi)}
\end{equation}
and the condition that $\nu_{21}(\xi(\mathbf{x}))>0$ requires that $\Gamma_{12}(\xi(\mathbf{x})) \Gamma_{21}(\xi(\mathbf{x})) > 0$. Note this guarantees a real-valued eigensystem (another way of defining hyperbolicity). 
To solve (\ref{eq:compaptHyper}), we look for a solution satisfying $\gamma(\mathbf{x}) = f(\xi(\mathbf{x}))$ for a function $f(\xi)$. This is inspired by the notion of a  \textit{simple wave} solution, which is of the  form $(\gamma(\mathbf{x}),\xi(\mathbf{x})) = \mathbf{f}(\theta(\mathbf{x}))$ for functions $\mathbf{f}(\theta)$ and $\theta(\mathbf{x})$ \cite[Ch.\;11.2.1]{evans10}. Here, we use  $\mathbf{f}(\theta)=(f(\theta),\theta)$. 

In order for the ansatz $\gamma(\mathbf{x}) = f(\xi(\mathbf{x}))$ to solve the PDE system (\ref{eq:compaptHyper}), we must require that 
\begin{equation}
\begin{aligned}\label{eq:reduceCompat}
f'(\xi(\mathbf{x})) \nabla \xi(\mathbf{x}) = \boldsymbol{\Gamma}(\xi(\mathbf{x})) \nabla \xi(\mathbf{x}).
\end{aligned}
\end{equation}
This condition requires $\nabla \xi(\mathbf{x})$ to be parallel to an  eigenvector $\mathbf{v}^{\pm}(\xi(\mathbf{x}))$. Then, $f'(\xi)$ is recovered from the corresponding eigenvalue.   Since the eigenvectors form an orthonormal basis, it is equivalent to solve 
\begin{equation}
\begin{aligned}
\begin{cases}\label{eq:methodCharac}
\Gamma_{12}(\xi(\mathbf{x})) \Gamma_{21}(\xi(\mathbf{x})) > 0,\\
\nabla \xi(\mathbf{x}) \cdot \mathbf{W} \mathbf{v}^{\sigma}(\xi(\mathbf{x})) = 0
\end{cases}
\end{aligned}
\end{equation}
with $\sigma= +$ or $-$ and with $\mathbf{W} = \mathbf{R}(\pi/2)$. For various boundary conditions,  (\ref{eq:methodCharac}) has a unique solution, which covers some or all of the specimen domain $\Omega$. This solution can be obtained analytically by the method of characteristics:  $\xi(\mathbf{x})$ is found to be constant along a family of straight \textit{characteristic lines}, each of which is perpendicular to $\mathbf{v}^{\sigma}(\xi(\mathbf{x}))$. If such a solution exists in only part of the domain $\tilde{\Omega} \subset \Omega$, we can attempt to cover the rest of $\Omega$ using other simple wave solutions (or other solutions altogether). In Fig.\;4(b), we use four simple wave solutions in the four corners of the sample, along with a constant actuation region where $\xi(\mathbf{x})\equiv \xi_0$ in the middle portion.

Before going on to describe the simple wave solutions behind Fig.\;4(b), we explain how to recover the rotation angle $\gamma(\mathbf{x})$. Suppose we have a simple wave solution per (\ref{eq:methodCharac}) on a sub-domain $\tilde{\Omega} \subset \Omega$ covered by characteristic lines.  Then, considering the equation (\ref{eq:reduceCompat}) and our ansatz $\gamma(\mathbf{x}) = f(\xi(\mathbf{x}))$, we find that 
\begin{equation}
\begin{aligned}\label{eq:gammaSolve}
\gamma(\mathbf{x}) = \int_0^{\xi(\mathbf{x})} \Gamma^{\sigma}(s) ds + \gamma_0, \quad \mathbf{x}\in\tilde{\Omega}
\end{aligned}
\end{equation}
for some real-valued constant $\gamma_0$. Since $\xi(\mathbf{x})$ remains constant along characteristic lines, so does $\gamma(\mathbf{x})$.

To obtain Fig.\;4(b), we first construct a simple wave solution to match the experiment in the upper left quadrant of the pattern.  We take this region to be the domain $(0,1)^2$ with the panels scaled appropriately on this domain without loss of generality.  The solutions on the other three quadrants are then obtained by mirror symmetry. Since the experiment appears to indicate a rarefaction fan-type shape, we parameterize our simple wave solutions using their boundary data along the line segment $\mathcal{L} := \{ s \mathbf{e}_2 \colon s \in (0,1)\}$, and consider data $\xi=\xi_{\text{b}}$ corresponding to a sharp increase in the opening angle at the origin.  Specifically, we use  $\xi_{\text{b}}(s) = \xi_0 \exp(-\lambda s)$, $s \in (0,1)$ for $\lambda \gg 1$ and $\xi_0 \in (0, 0.235 \pi)$. This choice gives  $\xi_{\text{b}}(0) = \xi_0$, $\xi_{\text{b}}(1)  \approx 0$, with a sharp increase in actuation near the origin.  We then set 
\begin{equation}
\begin{aligned}
\xi\big( s \mathbf{e}_2 + t \mathbf{W} \mathbf{v}^{+}( \xi_{\text{b}}(s)) \big) = \xi_{\text{b}}(s) \quad \text{ for } \quad   s \mathbf{e}_2 + t \mathbf{W} \mathbf{v}^{+}( \xi_{\text{b}}(s)) \in (0,1)^2
\end{aligned}
\end{equation}
to obtain our simple wave solution. This prescription solves (\ref{eq:methodCharac}) with characteristic lines that sweep from left to right throughout the domain. They start from a line parallel to $\mathbf{e}_2$, and tilt towards a line parallel to $\mathbf{e}_1$ for a value of $\xi_0 = 0.235\pi$.  The value of $\xi_0 = 0.11 \pi$ matches the bulk deformation inside the (essentially uniform) middle diamond region.  This value sets the rightmost characteristic, which is parallel to $\mathbf{W} \mathbf{v}^{+}( \xi_{\text{b}}(0) = 0.11\pi)  \approx 0.7 \mathbf{e}_1 + \mathbf{e}_2$. We  choose $\lambda = 10$  for the simulated solution; other large values of $\lambda$ do not alter the solution noticeably.  Again, we apply $\xi(\mathbf{x}) \equiv \xi_0 = 0.11\pi$ on the regions of $(0,1)^2$ not described by the simple wave solution and its mirror reflections. 
Having constructed $\xi(\mathbf{x})$, we find $\gamma(\mathbf{x})$ using (\ref{eq:gammaSolve}) with $\gamma_0 = -\int_0^{\xi_b(0)} \Gamma^{+}(s) ds$. This prescription solves (\ref{eq:compaptHyper}) on $(0,1)^2$  and mirroring the result  produces a solution to the full compatibility condition (\ref{eq:compaptHyper}) for the example in Fig.\;4(b). We plot the result using the numerical method discussed in Section \ref{s:Simulations}.

\subsection{Conformal map solutions}
We now consider rhombi-slit kirigami under the assumption that $\alpha = - \beta$.  From the formula for $\mathbf{A}(\xi)$ in the main text, the effective PDE reduces in this setting to
\begin{equation}
    \begin{aligned}\label{eq:conformalCase}
        \nabla \mathbf{y}_{\text{eff}}(\mathbf{x}) = \mu(\xi(\mathbf{x}))  \mathbf{R}(\gamma(\mathbf{x})) 
    \end{aligned}
\end{equation}
where $\mu_1(\xi) = \mu_2(\xi) = \mu(\xi) = \cos \xi - \alpha \sin \xi$.
Instead of using the compatibility condition to determine the angles $(\gamma(\mathbf{x}), \xi(\mathbf{x}))$, we simply observe that such effective deformations are conformal maps in the plane. So, they can be found using complex analysis.

Let $z = x_ 1 + i x_2$ denote a complex number and consider any complex analytic function  $f(z)$  on a domain $\Omega$ in the complex plane.  Then  $\mathbf{y}_{\text{c}}(\mathbf{x}) = \text{Re}[f(z)] \mathbf{e}_1 + \text{Im}[f(z)] \mathbf{e}_2$ has $\nabla \mathbf{y}_{\text{c}}(\mathbf{x}) = \mu_{\text{c}}(\mathbf{x}) \mathbf{R}(\gamma(\mathbf{x}))$ for a 2D rotation $\mathbf{R}(\gamma(\mathbf{x}))$ and dilatation $\mu_{\text{c}}(\mathbf{x})$. It is an effective deformation obeying (\ref{eq:conformalCase}) if we can match  $\mu_c(\mathbf{x})$ to an angle $\xi(\mathbf{x})$ such that $\mu(\xi(\mathbf{x})) = \mu_c(\mathbf{x})$.  This is  possible as long as $\mu_c(\mathbf{x})$  does not stray too far from $1$ (how far exactly depends only on the choice of $\alpha = -\beta$).

\textit{Optimization framework for conformal kirigami\;--\;}We generate conformal deformations $\mathbf{y}_{c}(\mathbf{x}; \boldsymbol{\tau}, \boldsymbol{\delta}, \boldsymbol{\kappa})$ as above, based on the family of complex polynomials of the form $f(z) = \sum_{k =1,\ldots, N} \tau_{k} (z - \delta_k  - i \kappa_k)^k$ with real valued parameters $\boldsymbol{\tau} = (\tau_1, \ldots, \tau_N)$, $\boldsymbol{\delta} = (\delta_1, \ldots, \delta_N)$ and $\boldsymbol{\kappa} = (\kappa_1, \ldots, \kappa_N)$.  To compare with the experiments, we notice that a proxy for boundary data is the list of 2D vectors $\bar{\mathbf{y}}_{1}, \ldots, \bar{\mathbf{y}}_{M}$ that represent the center points of each boundary panel in a deformed sample. Prior to deformation, these panels are also described by an analogous list of 2D vectors  $\bar{\mathbf{x}}_1, \ldots, \bar{\mathbf{x}}_M$. After extracting these lists, we optimize 
\begin{equation}
\begin{aligned}
\min_{\boldsymbol{\tau}, \boldsymbol{\delta}, \boldsymbol{\kappa}  \in \mathbb{R}^{N}} \sum_{m = 1}^{M} | \mathbf{y}_{\text{c}}(\bar{\mathbf{x}}_m; \boldsymbol{\tau},\boldsymbol{\delta}, \boldsymbol{\kappa}) -  \bar{\mathbf{y}}_{m}|^2
\end{aligned}
\end{equation}
 using \verb+fminunc+ in MATLAB. From the optimal $\mathbf{y}_{\text{eff}}(\mathbf{x}) =  \mathbf{y}_{\text{c}}(\mathbf{x}; \boldsymbol{\tau}_{\text{opt}}, \boldsymbol{\delta}_{\text{opt}}, \boldsymbol{\kappa}_{\text{opt}})$, we obtain the angles $\xi(\mathbf{x})$ and $\gamma(\mathbf{x})$. Finally, with $\mathbf{y}_{\text{eff}}(\mathbf{x}), \gamma(\mathbf{x}), \xi(\mathbf{x})$ known, we plot the results using the numerical method described in Section \ref{s:Simulations}.

\begin{figure}[h!]
\centering
\includegraphics[width =\linewidth]{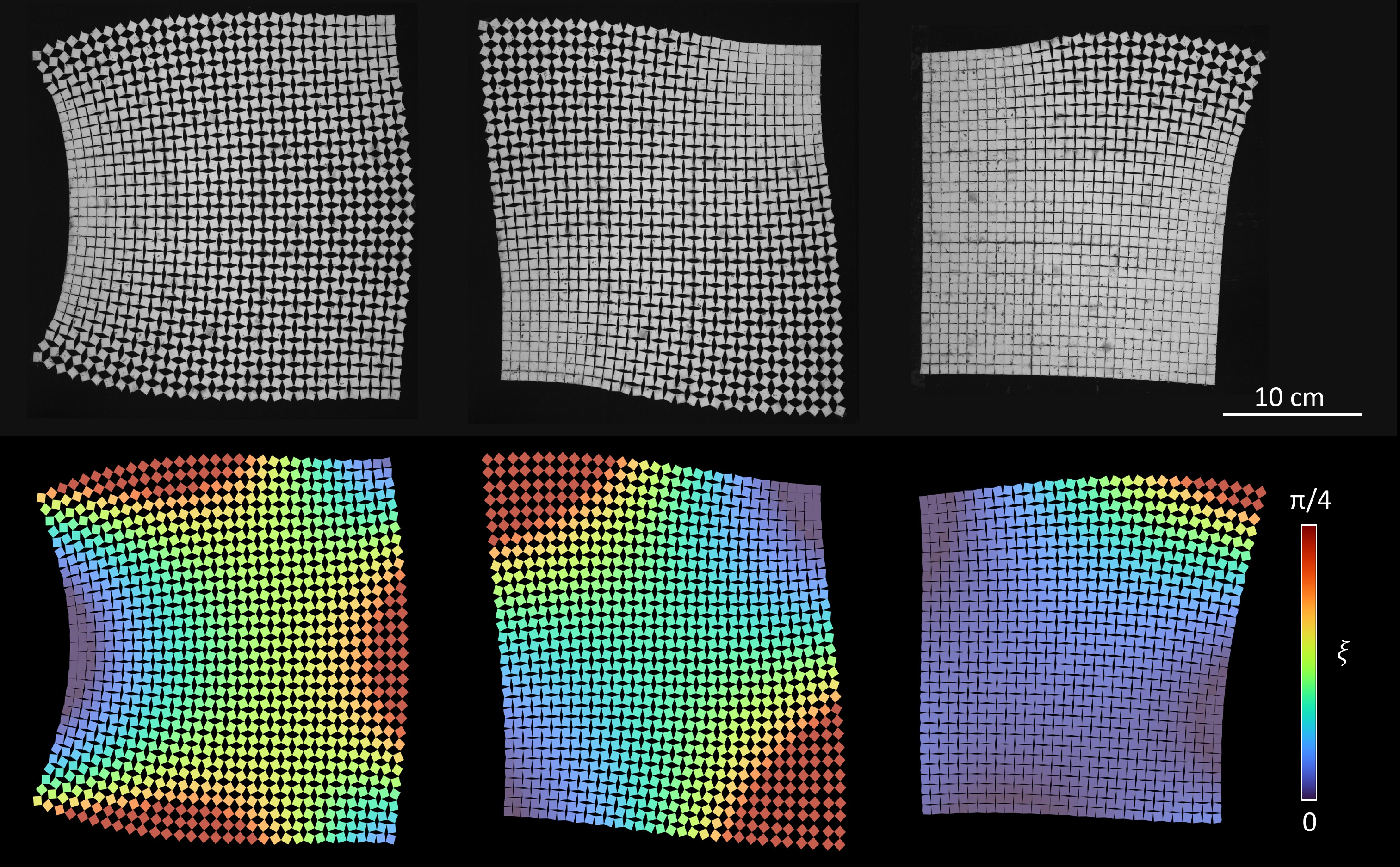}
\caption{Comparison between experiments (top) and simulations (bottom) for conformal kirigami.}
\label{fig:Conformal}
\end{figure}

This optimization procedure is extremely versatile. We used it to produce the simulation in Fig.\;4(e), based on the extracted boundary data from the RS center-pulling experiment (same figure). We have also done a number of other comparisons of different boundary data for the RS example. Fig.\;\ref{fig:Conformal} shows three additional comparisons between theory and experiment using the optimization procedure. All simulations in the conformal case were carried out with $N = 5$ ($15$ parameters are optimized) with the reference configuration  $\boldsymbol{\tau^0} = (1,0,0,0,0)$, $\boldsymbol{\delta^0} = \boldsymbol{0}$ and $\boldsymbol{\kappa^0} = \boldsymbol{0}$ chosen as the input data to the optimization.

\subsection{One-dimensional solutions}
Here we construct some universal solutions for both hyperbolic and elliptic rhombi-slit kirigami, via the one-dimensional ansatz $\xi(\mathbf{x}) = f(\mathbf{x} \cdot \mathbf{t})$ with a unit vector $\mathbf{t}$. The annular solutions in Fig.\;4(c) and (f) use the results of this section.

Enforcing the usual compatibility conditions (\ref{eq:compaptHyper}), we find for $\xi(\mathbf{x}) = f(\mathbf{x} \cdot \mathbf{t})$ that \begin{equation}
\begin{aligned}
\nabla \gamma(\mathbf{x}) = f'(\mathbf{x} \cdot \mathbf{t}) \Gamma_{12}(f(\mathbf{x} \cdot \mathbf{t})) (\mathbf{t} \cdot \mathbf{e}_2) \mathbf{e}_1 +  f'(\mathbf{x} \cdot \mathbf{t}) \Gamma_{21}(f(\mathbf{x} \cdot \mathbf{t})) (\mathbf{t} \cdot \mathbf{e}_1) \mathbf{e}_2.
\end{aligned}
\end{equation}
Since the partial derivatives $\partial_1 \partial_2 \gamma(\mathbf{x})$ and $\partial_2 \partial_1 \gamma(\mathbf{x})$ commute, we obtain that
\begin{equation}
\begin{aligned}\label{eq:ODEformula}
\Big( f'(s) \Gamma_{12}(f(s)) (\mathbf{t} \cdot \mathbf{e}_2)^2  - f'(s) \Gamma_{21}(f(s))  (\mathbf{t} \cdot \mathbf{e}_1)^2 \Big)' &= 0 \\
\Rightarrow \quad f'(s)\Big(  \Gamma_{12}(f(s)) (\mathbf{t} \cdot \mathbf{e}_2)^2 - \Gamma_{21}(f(s))  (\mathbf{t} \cdot \mathbf{e}_1)^2\Big) &= c_0
\end{aligned}
\end{equation}
for some constant $c_0 \in \mathbb{R}$. This last equation is an ODE in $f(s)$ that can be solved for a suitable initial condition $f(0)$. 

Taking one such solution, we recover its $\gamma(\mathbf{x})$ as follows. If $\mathbf{t} \cdot \mathbf{e}_2 = 0$, then  $\nabla \gamma(\mathbf{x}) = -c_0 \mathbf{e}_2$, yielding $\gamma(\mathbf{x}) = - c_0 (\mathbf{x} \cdot \mathbf{e}_2)+ d_0$ for some constant $d_0$.  Otherwise,
\begin{equation}
\begin{aligned}
\nabla \gamma(\mathbf{x}) =  \frac{c_0}{(\mathbf{t} \cdot \mathbf{e}_2)} \mathbf{e}_1  + f'(\mathbf{x} \cdot \mathbf{t})  \Gamma_{21}(f(\mathbf{x} \cdot \mathbf{t})) \frac{(\mathbf{t} \cdot \mathbf{e}_1)}{(\mathbf{t} \cdot \mathbf{e}_2)} \mathbf{t}
\end{aligned}
\end{equation}
and hence
\begin{equation}
\begin{aligned}
\gamma(\mathbf{x}) = \frac{c_0}{(\mathbf{t} \cdot \mathbf{e}_2)} (\mathbf{x} \cdot \mathbf{e}_1) +  \frac{(\mathbf{t} \cdot \mathbf{e}_1)}{(\mathbf{t} \cdot \mathbf{e}_2)}  \int_0^{f(\mathbf{x} \cdot \mathbf{t})} \Gamma_{21}(s) ds  + d_0
\end{aligned}
\end{equation}
for some constant $d_0$.  In either case, we obtain an expression for $\gamma(\mathbf{x})$.  

The annular solutions in the main text use $\mathbf{t} = \mathbf{e}_1$ or $\mathbf{e}_2$. We solve the ODE in (\ref{eq:ODEformula}) with $f(0) = 0$, and select the constant $c_0$ such that the resulting $f(s)$ is monotonically increasing on the interval $(0,1)$.

\section{Specimen fabrication and data extraction}
Our kirigami specimens are cut out of 1.5 mm-thick natural rubber sheets (McMaster-Carr 8633K71) using an 80 Watt Epilog Fusion Pro 32 laser cutter. To avoid burning the specimens and to produce clean cuts, the laser cutter is focused on the bottom face of the rubber sheet. The specimens are painted with white primer paint in order to facilitate the image processing procedure and to create high contrast with the black background.

\begin{figure}[h]
\centering
\includegraphics[scale = 1]{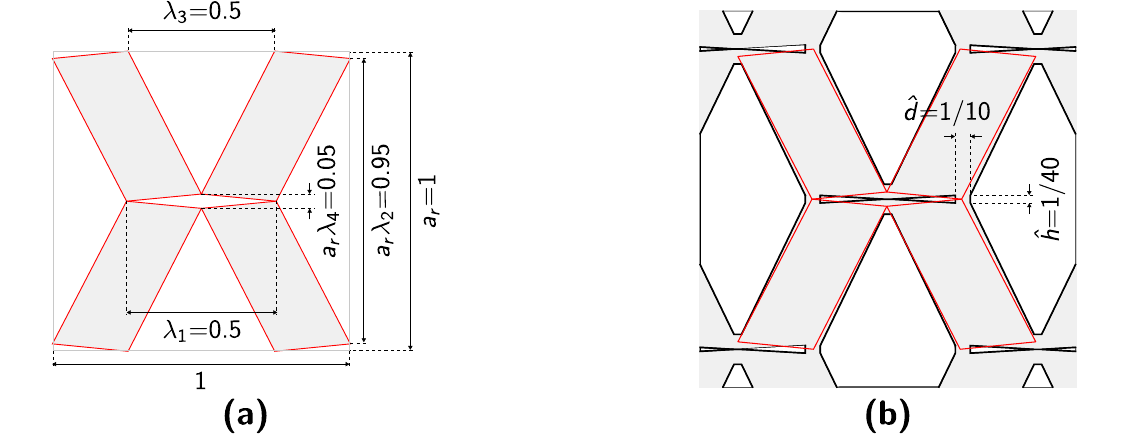}
\caption{(a) Idealized unit cell for the hyperbolic sample, with its characteristic dimensions. (b) A portion of the pattern used to fabricate an analog to (a); the thin red lines overlaid to the pattern represent the idealized unit cell. The only difference between (b) and (a) is the addition of flexure-like hinges of non-dimensional width $\hat{d}$ and length $\hat{h}$.}
\label{fig:Fab}
\end{figure}

The specific kirigami patterns discussed in the main article need to undergo slight modifications before being processed by the laser cutter. In particular, ideal point-like joints are replaced with compliant flexure-like hinges. This modification to the design is illustrated in Fig.~\ref{fig:Fab} for the hyperbolic sample. Fig.~\ref{fig:Fab}(a) represents the idealized pattern and Fig.~\ref{fig:Fab}(b) represents the pattern that is sent to the laser cutter.
In Fig.~\ref{fig:Fab}(b), we also indicate the normalized dimensions of the hinges: $\hat{d}=1/10$ and $\hat{h}=1/40$ as shown. 

The images of the samples are recorded by means of a FLIR 5-megapixel 35 fps camera with Edmund Optics lenses. The samples are anchored to the supporting plate via pins or double-sided tape. Quantitative information on the deformation is extracted through digital image processing in MATLAB. After converting the images to binary (using the \verb+imbinarize+ function), we obtain the centroid $\textbf{c}_i$, semi-major axis $\textbf{a}_i$, and semi-minor axis $\textbf{b}_i$ of a generic slit $i$ using the \verb+regionprops+ function. Then, we calculate $\xi_i$ from $\tan{\xi_i}=a_i/b_i$ and take $\gamma_i$ as the inclination of the major axis with respect to the horizontal; since we know the dimensions of the panels in the sample, we plot an idealized unit cell, deformed by an angle $\xi_i$ and rotated by $\gamma_i$, centered at the slit centroid $\textbf{c}_i$. We repeat this process for each slit. In Fig.\;4 of the main text, panels are only superimposed onto the right-half of each specimen and are colored according to the calculated value of $\xi_i$.

\bibliographystyle{apsrev4-1} 
\bibliography{bib}